\documentclass{aa}
\usepackage{graphics}
\begin{document}

%
\title{Optically bright Active Galactic Nuclei in the ROSAT-Faint Source Catalogue\thanks
{Tables 2, 3 and 4 are only available in electronic form at http://www.edpsciences.org}}

\author{M.-P. V\'eron-Cetty \inst{1}, S.K. Balayan \inst{2}, A. M. Mickaelian \inst{2}, 
R. Mujica \inst{3}, V. Chavushyan \inst{3}, S. A. Hakopian \inst{2}, D. Engels \inst{4}, 
P. V\'eron \inst{1}, F.-J. Zickgraf \inst{4}, W. Voges \inst{5}, D.-W. Xu \inst{6,5}
}


\institute{Observatoire de Haute Provence, CNRS, F-04870 Saint-Michel 
l'Observatoire, France \\ 
\email{mira.veron@oamp.fr; philippe.veron@oamp.fr} 
\and Byurakan Astrophysical Observatory and Isaac Newton Institute of Chile, 
Armenian branch, Byurakan 378433, Aragatzotn province, Armenia \\
\email{bal@moon.yerphi.am; aregmick@apaven.am; sanna@eugene.yerphi.am}
\and INAOE, Apdo. postal 51 y 216, 72000 Puebla, Pue., Mexico \\
\email{rmujica@inaoep.mx; vahram@inaoep.mx}
\and Hamburger Sternwarte, Gojenbergsweg 112, 21029 Hamburg, Germany \\
\email{dengels@hs.uni-hamburg.de; fzickgraf@hs.uni-hamburg.de}
\and Max-Planck-Institute f\"{u}r extraterrestrische Physik, Postfach 1312, 
D-85741, Garching, Germany \\
\email{whv@mpe-garching.mpg.de}
\and National astronomical observatories, Beijing 100012, China \\
\email{dwxu@bao.ac.cn}}

\titlerunning{Bright AGN in the ROSAT-FSC}
\authorrunning{V\'eron-Cetty et al.}

\date{Received 10 June 2003/Accepted 17 October 2003}

\abstract{To build a large, optically bright, X-ray selected AGN sample we have 
correlated the ROSAT-FSC catalogue of X-ray sources with the USNO catalogue limited 
to objects brighter than O=16.5 and then with the APS database. Each 
of the 3\,212 coincidences was classified using the slitless Hamburg spectra. 493 
objects were found to be extended and 2\,719 starlike. Using both the extended 
objects and the galaxies known from published catalogues we built up a sample of 
185 galaxies with O$_{\rm APS}$$<$17.0 mag, which are high-probability counterparts 
of RASS-FSC X-ray sources. 130 galaxies have a redshift from the literature and 
for another 34 we obtained new spectra. The fraction of Seyfert galaxies in 
this sample is 20\%. To select a corresponding sample of 144 high-probability 
counterparts among the starlike sources we searched for very blue objects in an 
APS-based color-magnitude diagram. Forty-one were already known AGN and for another 
91 objects we obtained new spectra, yielding 42 new AGN, increasing their number 
in the sample to 83. This confirms that surveys of bright QSOs are still significantly 
incomplete. On the other hand we find that, at a flux limit of 0.02 count s$^{-1}$ 
and at this magnitude, only 40\% of all QSOs are detected by ROSAT.
}
\maketitle

\today


\section{Introduction}

 The ROSAT All-Sky Survey Bright Source Catalogue (RASS-BSC) is derived from
the all-sky survey performed during the ROSAT mission in the energy band 
0.1-2.4 keV; it contains 18\,811 sources down to a limiting ROSAT-PSPC 
count-rate of 0.05 count s$^{-1}$ (Voges et al. \cite{voges99}). The 2\,012 
brightest (count-rate above 0.2 count s$^{-1}$) high galactic latitude 
($\vert$b$\vert$$>$30$^{\rm o}$) sources of the BSC catalogue have been 
tentatively optically identified by Schwope et al. (\cite{schwope00}). The ROSAT 
Faint Source Catalogue (FSC)(Voges et al. \cite{voges00}) contains 105\,924 
sources and represents the faint extension of the RASS-BSC.

 The Hamburg QSO Survey is a wide-angle objective prism survey for finding 
bright QSOs in the northern sky. The survey plates have been taken with the 
former Hamburg Schmidt telescope now located at the Spanish-German Center in 
Calar Alto (Spain). A 1$\fdg$7 objective prism has been used providing 
unwidened spectra with a dispersion of 1\,390 \AA\ mm$^{-1}$ at H$\gamma$
(Hagen et al. \cite{hagen95}). The slitless spectra allow the classification 
of objects brighter than about B=17.0. The first study of a sample of 
previously known QSOs with the ${\it Einstein\,Observatory}$ has shown that, 
as a class, they are luminous X-ray 
emitters (Tananbaum et al. \cite{tananbaum79}). Bade et al. (\cite{bade98}) and
Zickgraf et al. (\cite{zickgraf03}) have described a way to identify X-ray 
sources by using the Hamburg survey plates; this process has been applied to 
a subsample of 5\,341 sources from the ROSAT-BSC. 
 
 In the present paper we try to find all AGN brighter than O=16.5 in a subarea of the 
ROSAT-FSC. As the FSC survey is only about 2.5 times deeper than the BSC survey, 
we do not expect a significant difference in the identification content of 
these two samples but, because the number of sources in the FSC survey is much 
larger, we expect the discovery of many additional optically bright AGN.
As the ROSAT survey is a low-energy survey,  
it contains a relatively high fraction of Narrow-Line Seyfert 1 Galaxies
(NLS1s) because they usually have a soft X-ray excess (see for instance
Laor et al. \cite{laor97}); we therefore expect to find new bright NLS1s. \\

 In the following we call AGN (or Active Galactic Nuclei) QSOs, BL~Lac 
objects, Seyfert galaxies and Liners. QSOs are defined as Seyfert 1 galaxies 
brighter than M$_{\rm B}$=--23.0\footnote{Throughout this 
work we used H$_{\rm o}$=50 km s$^{-1}$ Mpc$^{-1}$ and q$_{\rm o}$=0} .
 
\section{Analysis}

\subsection{The sample}

 In the area of the sky defined by $\delta>$ 0$^{\rm o}$ and $\vert$b$\vert>$ 
30$^{\rm o}$ (10\,313 deg$^2$) there are 29\,321 ROSAT-FSC sources. 

\begin{figure}[ht]

\resizebox{8.8cm}{!}{\includegraphics{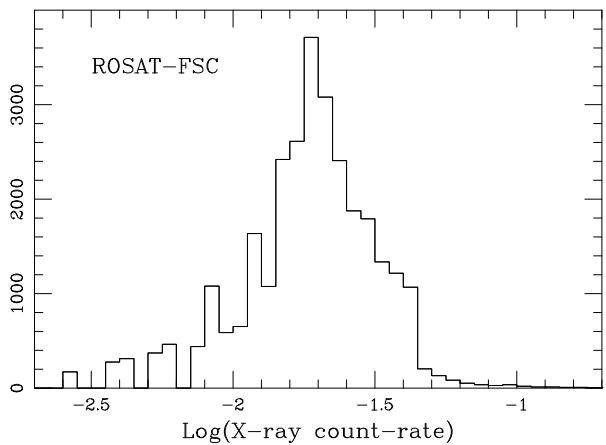}}

\caption{\label{histo_fx} Histogram of the logarithm of the X-ray count rates 
(in counts s$^{-1}$) of the 29\,321 ROSAT-FSC sources.}

\end{figure}

 Figure \ref {histo_fx} shows the histogram of the logarithm of the X-ray count 
rates of these sources (in counts s$^{-1}$). Below 0.02 counts s$^{-1}$ the 
survey is obviously quite incomplete, reflecting the fact that its sensitivity 
limit is not uniform over the sky. On the other hand 130 sources 
have a count rate greater than 0.05 counts s$^{-1}$ but were not included 
in the BSC as their detected photon number was less than 15. They were 
included in the FSC. 

\begin{figure}[ht]

\resizebox{8.8cm}{!}{\includegraphics{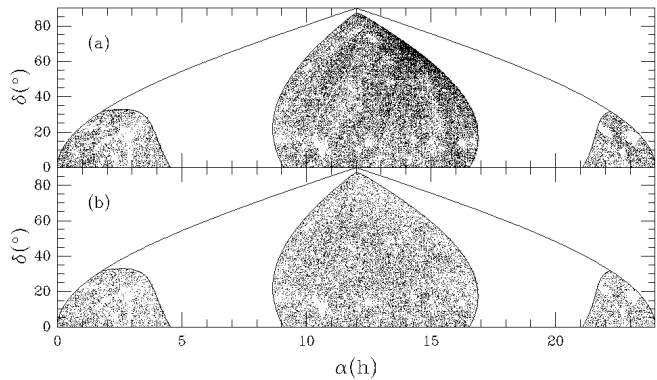}}

\caption{\label{carte} Distribution over the sky of (a) the 29\,321 ROSAT-FSC 
sources, (b) the 15\,848 sources brighter than 0.02 counts s$^{-1}$.}

\end{figure}

 Figure \ref {carte} shows the distribution over the sky of (a) the 29\,321 
sources and (b) the 15\,848 sources with F$_{\rm x}>0.02$ counts s$^{-1}$. 
The second distribution is much more uniform than the first and, 
in the following, we shall restrict our analysis to this subsample. \\

\subsection{Identification methodology}

  Our aim being to identify all bright extragalactic objects associated with a ROSAT-FSC
source in the area defined above, we chose to make use of the Hamburg slitless survey
of the northern sky which, in principle, allows us to determine the  nature of every 
object brighter than O$\sim$17.0. However this would require us to visually inspect all objects 
within a given radius around each of the 15\,848 X-ray sources in the sample. This is 
obviously not feasible. We therefore decided to preselect all bright objects lying near 
the X-ray positions in the APS database, which is a catalogue of all objects visible on the 
Palomar Sky Survey plates, including magnitude, colour and classification as starlike or 
extended. But this could not be easily done as the APS was not directly available 
(when we started this work). The USNO catalogue is similar to the APS catalogue except that
it does not classify the listed objects as starlike or extended; but it is available on 
CD-ROMs. It was therefore easy to select all bright USNO objects near the X-ray positions. 
We then sent the list of 3\,776 selected objects to the University of Minnesota where 
a batch job was run to find their APS magnitude, colour and classification. 

  However, as very bright stars (O$<$12.0) are saturated on the DSS1 images and bright 
extended galaxies are poorly recognized by the automatic extraction procedures of both 
the USNO and APS databases, to find these objects we had to cross-correlate the X-ray 
catalogue with various catalogues of bright stars and galaxies. We ended up with 3\,212 
objects found in the APS database, 685 bright stars and 91 additional galaxies.
 
  The APS classification as ''star-like" or ''extended" were then evaluated by 
visual screening of the DSS2 images. 2\,719 APS objects were found to be 
''star-like" and 493 ''extended". All 3\,212 APS objects were subsequently 
classified using the digitized objective prism spectra of the Hamburg Quasar 
Survey.

  Follow-up observations were started for the ''extended" objects and additional 
galaxies having O$_{\rm APS}$$<$17.0, showing the sample to be a mixture of 
galaxies containing 20\% AGN.

  The ''star-like" objects were separated into several subsamples according to 
their APS colour and magnitude. A sample of 144 high-probability AGN candidates 
was selected. Based on literature data and our own follow-up spectroscopy, 83 
of them are now confirmed AGN.
  
\subsection{The USNO and APS databases}

 To identify the X-ray sources with relatively bright optical counterparts we have
cross-correlated our ROSAT sample with the USNO-A2.0 catalogue (Monet et al. 1996)
from which we have extracted all objects in the magnitude range O=12.0-16.5 
(4\,420\,441 {\it i.e.} 428 deg$^{-2}$). We have found 25\,549 objects located 
within 2$\arcmin$ of one of the 15\,848 X-ray sources.

\begin{figure}[ht]

\resizebox{3.0cm}{!}{\includegraphics{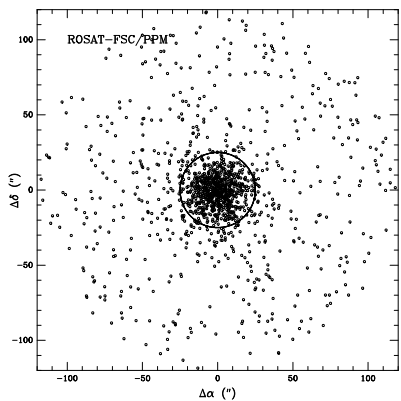}}

\caption{\label{rosat_ppm} Position differences for the coincidences between 
ROSAT-FSC X-ray sources and PPM stars within two arc minutes. The radius of the 
circle is 25$\arcsec$.}

\end{figure}

 As the USNO catalogue positions are rather poor for very bright objects, we have 
extracted the 1\,306 PPM (Positions and Proper Motions) stars (R\"oser 
\& Bastian \cite{roser}) located within 120$\arcsec$ of a ROSAT position. The PPM
star catalogue contains 181\,731 stars north of declination --2$\fdg$5 , brighter 
than about m$_{pg}$=11.0. Fig. \ref{rosat_ppm} shows the distribution of the 
position differences between the ROSAT and PPM positions. The circle drawn on the 
figure has a radius of 25$\arcsec$. The number of coincidences within this circle 
is 704 (with 19 X-ray sources being associated with two PPM stars), while the 
expected number of chance coincidences is 27. There are therefore 685 X-ray sources 
within 25$\arcsec$ of at least one PPM star. We consider these 685 sources 
as identified; we ignored them in the following and we are left with 15\,163 X-ray 
sources. 

\begin{figure}[ht]

\resizebox{8.8cm}{!}{\includegraphics{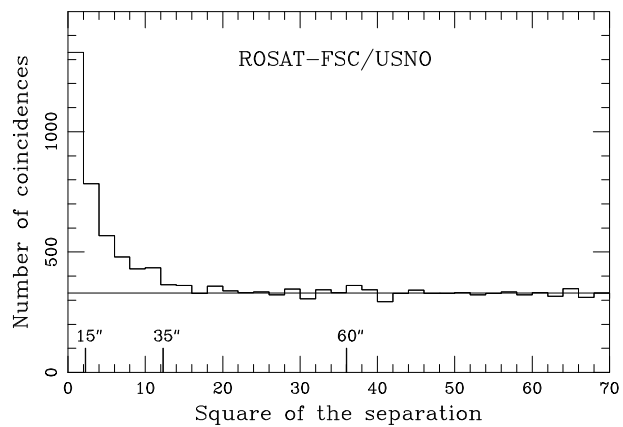}}

\caption{\label{histo} Histogram of the square of the separation between the X-ray 
and USNO positions for the 25\,549 associations in units of 100 square arcsec.}

\end{figure}

 The histogram of the separation between the optical and X-ray positions (Fig.
\ref{histo}) shows that, for separations larger than 35$\arcsec$, there is an 
overwhelming majority of chance coincidences. For smaller separations, the 
fraction of real associations rapidly increases. This empirically defined 
limit is a compromise to maximize the number of real associations and minimize 
the number of chance coincidences. It is not straightforward to determine the 
fraction of real associations having a separation larger than 35$\arcsec$ and 
therefore lost by using this limit. For separations smaller 
than 35$\arcsec$, there are 3\,776 coincidences, half being expected by chance 
and half real. Including the PPM stars, about 17\% of the X-ray FSC sources are 
therefore physically associated with a relatively bright object.

 Zickgraf et al. (\cite{zickgraf03}) found that the 90\% error radius for the 
X-ray positions in the BSC is 21$\arcsec$. This is significantly smaller than 
the value of 35$\arcsec$ we used for the separation limit for real associations.
This difference can probably be ascribed to the differences in count rates 
between the two samples.

 In total there are 3\,364 X-ray sources with at least one USNO object within 
35$\arcsec$: 2\,985 with a single USNO object, 348 with two, 29 with three and 
2 with four. \\

 In the APS database (Cabanela et al. \cite{cabanela03}), the objects are 
classified as starlike or resolved (galaxies). The photometric calibration for the 
starlike objects uses a magnitude-diameter relation derived from photoelectric 
calibrating sequences. The O magnitudes have a mean r.m.s. of 0.15-0.20 mag over 
the range 14-20 mag. These magnitudes are not as reliable for objects brighter 
than 12$^{\rm th}$ mag because of the diffraction pattern. For objects brighter 
than 8$^{\rm th}$ mag, photometry is not available. For galaxies, the integrated 
magnitudes are obtained from a density-to-intensity conversion. APS derived 
galaxy magnitudes show no systematic photometric errors and a typical r.m.s. 
scatter of 0.2 to 0.3 magnitudes. The O$_{\rm APS}$ and B magnitudes for stars 
are equal on average with a dispersion of 0.26 mag (Mickaelian et al. 
\cite{mickaelian99}). \\

 In contrast, the USNO catalogue makes no distinction between starlike and extended 
objects. As a consequence, the magnitudes derived for galaxies are unreliable. 
The typical photometric error for starlike objects is about 0.31 mag r.m.s. (Mickaelian 
et al. \cite{mickaelian01}). Bright objects tend to saturate. The magnitudes 
reported for such objects are generally too high. \\

\begin{figure}[ht]

\resizebox{8.8cm}{!}{\includegraphics{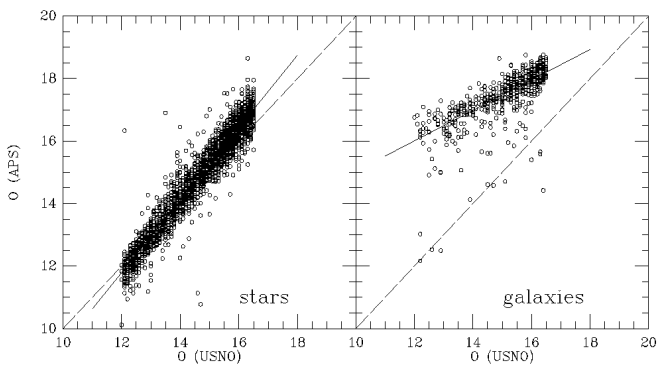}}

\caption{\label{mag_mag} Comparison of the USNO and APS O magnitudes for starlike
objects (left panel) and for galaxies (right panel). The solid lines represent the 
best linear fits between the two sets of data; the dotted lines have a slope of unity.}

\end{figure}

 Figure \ref{mag_mag} shows the comparison of the USNO and APS O magnitudes for starlike 
objects (left panel) and for galaxies (right panel). While the agreement between the
two sets of magnitudes is reasonably good for starlike objects, it is very poor for 
galaxies. \\

\begin{figure}[ht]

\resizebox{8.8cm}{!}{\includegraphics{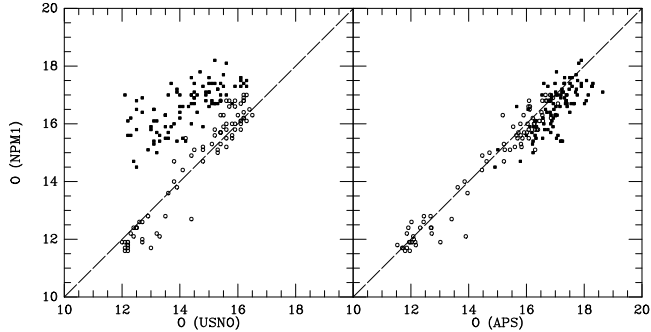}}

\caption{\label{comp_mag} Comparison of the USNO and APS O magnitudes with the NPM1 O
magnitudes for 95 galaxies (filled squares) and 81 stars (open circles). The dotted 
lines have a slope of unity.}

\end{figure}

 Figure \ref{comp_mag} shows the comparison of the USNO and APS O magnitudes 
with the NPM1 O magnitudes (Klemola et al. \cite{klemola87}) for 95 galaxies and 
81 stars.
 The least square solutions for galaxies are: \\
        O$_{\rm NPM1G}$ = 0.47$\times$O$_{\rm USNO}$ + 11.3  ($\sigma$ = 0.54 mag) \\
        O$_{\rm NPM1G}$ = 0.90$\times$O$_{\rm APS}$  -- 0.29  ($\sigma$ = 0.50 mag) \\

 This confirms that, for galaxies, the APS magnitudes are more reliable than the USNO 
magnitudes. We therefore extracted from the APS database objects within 10$\arcsec$ 
from the 3\,776 USNO objects, brighter than O=19.0. A number of these have no APS 
counterpart because the required plates were not available and 49 because there was 
no match. We checked these 49 USNO objects on the DSS2 images (Lasker et al 
\cite{lasker96}) and found that all of them, except one, have a starlike appearance. 
There are 3\,212 coincidences between USNO and APS objects (the 3\,212 sample), 
2\,652 with starlike and 560 with extended objects. 

 As the APS classification based on DSS1 is not always correct, all 3\,212 images have 
been checked on the DSS2 plates. Of these, 2\,719 were classified as 
starlike and 493 as extended. Out of the 560 APS galaxies, 135 have been reclassified 
as starlike; 45 of them appear as binaries so that in the APS they appear as extended. 
On the other hand, 68 APS starlike objects have an extended image on the DSS2  

\subsection{The Hamburg database}

 All 3\,212 objects have been tentatively classified using the Hamburg slitless 
plates on the basis of a classification scheme introduced during the ROSAT-BSC 
identification program (Bade et al. \cite{bade98}). The statistics of this
classification is given in Table \ref{hamburg}. Of the 493 objects classified as 
extended, based on direct images of the DSS2 plates, four could not be classified
spectroscopically because of overlapping spectra. Based on their continua the
489 remaining objects were classified as ``galaxy" (red continuum),``blue galaxy" 
(blue continuum) or ``AGN" (prominent UV excess). Sixty one of the galaxies are
member of a pair (23) or of a group (38). Among the starlike objects,
2\,181 were classified as stars due to absorption features in their 
spectra. The bright saturated objects (O$_{\rm APS}<$12.0) were also considered 
as stars. Moreover 383 starlike objects, too weak for the study of absorption 
features, were classified on the basis of their continuum: red (RED-WK), blue 
(BLUE-WK) and extremely blue (EBL-WK). These objects have typically 
O$_{\rm APS}>$16.0. Twenty two starlike objects were classified as QSOs due to 
the presence of a prominent UV excess and/or of emission lines. Fifty five 
objects could not be classified because of overlapping spectra and one more 
because its spectrum did not fit the classification scheme. 

\begin{table}
\caption{\label{hamburg} Statistics of the HQS type of 3\,212 optical associations 
with 2\,982 ROSAT FSC sources.}
\begin{center}
\begin{tabular}{lrr}
\hline
HQS types   &    Number &   \%  \\
\hline
AGN         &         6 &   0.2 \\ 
Blue galaxy &       215 &   6.7 \\  
Galaxy      &       268 &   8.3 \\
QSO         &        22 &   0.7 \\     
EBL-WK      &        84 &   2.6 \\  
BLUE-WK     &       212 &   6.6 \\  
RED-WK      &        87 &   2.8 \\  
W-Dwarf     &        20 &   0.6 \\ 
Star-BA     &       137 &   4.3 \\ 
Star-FG     &      1325 &  41.3 \\ 
Star-K      &       674 &  21.0 \\ 
Star-M      &        25 &   0.8 \\  
Saturated   &        81 &   2.5 \\ 
Overlap     &        55 &   1.7 \\ 
Unidentified&         1 &   0.0 \\ 
\hline 
Total       &      3212 & 100.0 \\
\hline
\end{tabular}
\end{center}
\end{table}
\normalsize

 The HQS classifications have been checked with already known objects. In our 
sample there are 107 confirmed QSOs. Two could not be classified with 
the Hamburg material because of overlapping spectra. Seventeen have been classified 
as QSOs, one as an AGN, 56 as EBL-WK, 22 as BLUE-WK, two as 
RED-WK (J15563+0309, a low luminosity QSO with M$_{\rm B}$=--23.2 and KUV\,16355+4146 
at z=0.765, with (O--E)=1.60) and seven as blue galaxies. These blue galaxies 
are in fact low luminosity QSOs (M$_{\rm B}$$>$--24.0). On the other hand eleven EBL-WK
objects have been shown to be stars, as well as 21 BLUE-WK and two RED-WK objects. 
Clearly the EBL-WK objects are good QSO candidates as well as the BLUE-WK, 
although to a lesser degree. When they are not QSOs, they usually turn out to be WDs 
or CVs.

\section{Galaxies}
 
 To find Seyfert galaxies we turned our attention to the 493 objects extended on
the DSS2,
found within 35$\arcsec$ of an FSC source. If there were no real association, 
the number of coincidences within 15$\arcsec$ would be 0.184 times the number 
of coincidences within 35$\arcsec$ (the ratio of the respective areas). On 
the other hand, if all associations were real, Fig. \ref{histo} shows that 
almost all identifications would be located within 35$\arcsec$ and about 50\% within 
15$\arcsec$. In a sample in which the fraction of coincidences within 15$\arcsec$ 
is $x$, the fraction $y$ of true associations would be $y$=($x$--0.184)/(0.5--0.184). 
218 (x=44) of the 493 galaxies are within 15$\arcsec$ of an X-ray source
which shows that a large fraction of them ($y$$\sim$80\%) are genuine associations.

  As we have seen above, sixty eight of the objects found to be extended on the 
DSS2 were classified as starlike in the APS data base. Three of them (NGC\,1085, 
IC\,2439 and KUG\,2323+85) were previously found to be bright galaxies and are 
included in Table \ref{galaxies}. The 65 others are probably compact galaxies; 
47 are classified as ''blue galaxies" on the slitless HQS spectra (in fact, ten 
of them are already known to be Seyfert 1s or low luminosity QSOs). 
 
 As the spectroscopic observations took place before we had the 
opportunity to check the APS classification on the DSS2 images (which is a rather 
tedious procedure as it is done visually rather than automatically), these 65 
objects were not included in our original list of galaxy candidates.

 Of the 493 extended objects, 155 are brighter than O$_{\rm APS}$=17.0.  
Ninety four of them were classified as galaxies
in the APS database. To these 94 galaxies we added 91 bright galaxies not 
appearing in the USNO or APS databases but found by cross-correlating the ROSAT 
FSC with catalogues of bright galaxies (NGC, UGC, NPM1G, Mark, Zw etc.). As the 
surface density in these catalogues is small, we accepted as identification all 
coincidences within 35$\arcsec$. These 185 galaxy candidates are listed in Table 
\ref{galaxies}. The redshift of 130 of them were known from the literature. 

\begin{table*}
\caption{\label{galaxies} List of the 185 galaxy candidates. Col. 1: right 
ascension, col. 2: declination, col. 3: separation between 
the ROSAT and USNO positions in arcsec, col. 4: X-ray count rate (count s$^{-1}$), col. 5:
logarithm of the 0.1-2.4 keV X-ray luminosity (in erg s$^{-1}$), col. 6: 
magnitude O$_{\rm USNO}$, col. 7: magnitude O$_{\rm APS}$, col. 8: redshift, col. 9: 
classification (abs: galaxies with absorption lines only; em: emission line galaxies,
but too weak to make a classification possible; C stands for Composite spectrum; 
S or S?: the object is certainly an AGN, but the lines are too weak for a more 
precise classification); col. 10: references: (1) 
BAO; (2) OHP, col. 11: name.}
\begin{center}
\begin{tabular}{rrr|rrr|r|cc|cc|c|l|l|l}
\hline \multicolumn{6}{c}{J2000 optical position} \\
\hline
h & m  & s & $^{\circ}$ & ' & " & sep(") & c/r & L$_{\rm X}$ & O$_{\rm US}$ & O$_{\rm APS}$& z & & ref. & Name  \\
\hline
  0 & 14 &  7.46 & 10 & 34 & 13.3 &   5.3 & 0.036 &     & 15.0 & 16.89 &       &       & &                   \\
  0 & 19 & 37.80 & 29 & 56 &  2.0 &  27.2 & 0.042 &42.02&      & 15.37 & 0.024 &       & & NGC   76          \\
  0 & 24 & 30.24 & 13 & 35 & 50.3 &   2.4 & 0.032 &     & 12.5 & 16.63 &       &       & & NPM1G+13.0013     \\
  0 & 46 & 29.20 &  8 & 25 & 59.6 &  18.9 & 0.023 &42.18& 15.5 & 16.32 & 0.039 &       & & UGC 482           \\
  0 & 54 & 45.23 & 16 & 26 & 17.2 &   9.2 & 0.038 &42.37& 13.6 & 16.56 & 0.038 &       & & MCG+03.03.008     \\
  0 & 54 & 48.64 & 28 & 52 &  1.1 &  26.7 & 0.024 &41.74&      & 15.66 & 0.023 &       & & UGC   555         \\
  0 & 57 & 48.89 & 30 & 21 &  8.8 &   7.3 & 0.043 &41.67&      & 13.14 & 0.016 & Liner & & NGC  315          \\
  1 & 16 & 14.83 & 31 &  2 &  1.8 &  34.2 & 0.021 &41.42&      &       & 0.017 &       & & NGC  452          \\
  1 & 20 & 38.97 & 29 & 41 & 43.9 &   9.4 & 0.031 &41.85& 16.0 & 15.40 & 0.023 &       & & IC 1672           \\
  1 & 30 & 35.46 & 19 & 36 & 31.0 &   4.2 & 0.034 &42.43& 13.7 & 16.35 & 0.043 &       & & UGC  1077         \\
  2 & 43 & 48.71 & 14 & 53 & 13.1 &  33.8 & 0.027 &42.39&      & 16.49 & 0.046 &       & & IC 1835           \\
  2 & 46 & 25.36 &  3 & 36 & 27.3 &   7.1 & 0.078 &42.25& 12.5 & 14.91 & 0.023 &       & & NGC 1085          \\
  2 & 51 & 12.98 & 13 & 11 & 31.3 &  16.1 & 0.031 &41.92& 13.8 & 16.34 & 0.025 &       & & UGC 2337B         \\
  2 & 58 & 51.14 &  6 & 22 & 25.9 &   5.4 & 0.026 &42.35& 13.5 & 16.88 & 0.045 &       & & LEDA 074274       \\
  7 & 50 &  8.28 & 55 & 23 &  2.9 &  18.6 & 0.032 &41.70& 12.6 & 15.43 & 0.019 & abs   & 2 & UGC 4035          \\
  8 &  1 & 55.04 & 62 & 32 & 15.0 &   6.0 & 0.034 &42.88& 14.0 &       & 0.072 &       & &                   \\
  8 & 30 & 29.14 & 69 &  1 & 42.8 &  12.3 & 0.022 &41.66& 13.5 &       & 0.022 &       & & Zw 331.048        \\
  8 & 36 & 45.79 & 48 & 42 &  0.7 &  30.1 & 0.027 &     & 13.7 & 16.24 &       &       & & MCG+08.16.019     \\
  8 & 39 & 55.44 & 74 &  5 & 15.9 &  32.2 & 0.024 &     & 15.1 &       &       &       & & NPM1G+74.0038     \\
  8 & 55 & 38.01 & 78 & 13 & 25.3 &   5.0 & 0.024 &40.41&      &       & 0.005 & Liner & & NGC 2655          \\
  9 &  6 & 44.71 &  3 & 36 &  0.9 &  28.1 & 0.023 &     & 14.2 &       &       &       & &                \\
  9 &  8 & 38.35 & 32 & 35 & 34.0 &  15.9 & 0.033 &41.44& 12.4 & 15.79 & 0.014 &       & & IC 2439           \\
  9 & 11 & 37.50 & 60 &  2 & 15.0 &  22.9 & 0.026 &40.45&      & 12.12 & 0.005 & S     & & NGC 2768          \\
  9 & 12 & 38.04 & 75 & 38 & 55.4 &  13.2 & 0.040 &42.81& 12.8 &       & 0.061 & H II  & 1,2& NPM1G+75.0038     \\
  9 & 16 &  3.95 & 17 & 37 & 43.6 &  10.1 & 0.060 &42.34&      & 14.63 & 0.029 &       & & NGC 2795          \\
  9 & 16 & 50.01 & 20 & 11 & 54.6 &  19.8 & 0.021 &41.85&      & 15.04 & 0.028 &       & & NGC 2804          \\
  9 & 19 & 27.31 & 33 & 47 & 26.4 &  26.4 & 0.029 &41.65& 13.7 & 16.83 & 0.019 &       & & LEDA 139185       \\
  9 & 20 &  2.10 &  1 &  2 & 18.0 &  15.9 & 0.046 &41.76&      & 14.27 & 0.017 &       & & UGC  4956         \\
  9 & 23 & 25.93 & 22 & 19 &  1.0 &  20.3 & 0.081 &42.50& 13.9 &       & 0.030 &       & & UGC  4991b        \\
  9 & 23 & 24.36 & 22 & 18 & 47.3 &   5.4 & 0.081 &42.60& 13.4 &       & 0.034 &       & & UGC  4991a        \\
  9 & 23 & 29.19 & 25 & 46 &  9.6 &  13.4 & 0.021 &42.35& 12.5 & 15.73 & 0.050 & S2    & 1,2 & Zw 121.071        \\
  9 & 24 & 14.29 & 49 & 15 & 14.9 &  32.9 & 0.021 &40.86&      & 14.71 & 0.009 &       & & NGC 2856          \\
  9 & 25 & 28.30 & 23 & 36 & 31.6 &   1.5 & 0.035 &42.21& 12.2 & 16.12 & 0.033 & em    & 1 & Zw 121.086        \\
  9 & 25 & 42.55 & 11 & 25 & 55.3 &  14.6 & 0.034 &41.25&      & 14.31 & 0.011 &       & & NGC 2872          \\
  9 & 32 & 10.13 & 21 & 30 &  4.4 &  18.1 & 0.030 &39.71&      &       & 0.002 & H II  & & NGC 2903          \\
  9 & 32 & 16.98 &  9 & 41 &  0.5 &  28.6 & 0.023 &     & 14.8 & 16.74 &       &       & & NPM1G+09.0188     \\
  9 & 38 & 12.30 &  7 & 43 & 39.8 &  19.9 & 0.033 &     & 13.8 & 16.84 &       &       & & Zw 035.017        \\
  9 & 38 & 32.90 & 17 &  1 & 52.0 &  17.9 & 0.040 &42.13&      & 16.00 & 0.028 &       & & NGC 2943          \\
  9 & 45 & 28.84 & 56 & 32 & 53.4 &   2.1 & 0.046 &43.58& 15.8 & 16.91 & 0.139 & abs   & 1 &                   \\
  9 & 50 & 21.61 & 72 & 16 & 44.1 &  22.7 & 0.025 &40.23&      &       & 0.004 & S1.9  & & NGC 2985          \\
  9 & 59 & 39.60 &  0 & 35 & 12.1 &  24.1 & 0.028 &42.72& 13.1 & 16.42 & 0.066 &       & & UGC  5370         \\
 10 &  1 & 57.80 & 55 & 40 & 47.1 &   4.2 & 0.049 &40.53&      & 12.23 & 0.004 & S2    & & NGC 3079          \\
 10 &  6 &  7.45 & 47 & 15 & 45.4 &  22.0 & 0.036 &41.98&      & 15.09 & 0.025 &       & & NGC 3111          \\
 10 & 13 & 50.50 & 38 & 45 & 53.8 &  27.8 & 0.028 &41.80&      & 14.15 & 0.023 &       & & NGC 3158          \\
 10 & 27 & 49.91 & 36 & 33 & 34.7 &  24.4 & 0.037 &42.34&      & 16.02 & 0.037 & S2    & & HS 1024+3648      \\
 10 & 32 & 33.34 & 65 &  2 &  0.9 &  33.2 & 0.023 &40.55&      & 14.42 & 0.006 &       & & NGC 3259          \\
 10 & 47 & 10.14 & 72 & 50 & 20.9 &   8.5 & 0.025 &40.94&      &       & 0.009 &       & & NGC 3348          \\
 10 & 51 & 29.91 & 46 & 44 & 41.5 &  24.1 & 0.026 &     & 13.3 & 16.00 &       &       & & NPM1G+47.0182     \\
 10 & 58 & 13.77 &  1 & 36 &  6.3 &   8.8 & 0.069 &42.63& 16.5 &       & 0.038 & abs   & 1 & UGC 6057c         \\
 10 & 58 & 13.04 &  1 & 36 & 24.3 &  29.9 & 0.069 &42.65& 15.1 &       & 0.039 &       & & UGC 6057b         \\
 11 &  0 &  5.35 & 14 & 50 & 26.5 &  28.4 & 0.026 &40.45& 15.9 &       & 0.005 &       & & NGC 3485          \\
 11 &  3 & 11.35 & 27 & 58 & 21.1 &   7.8 & 0.034 &40.56& 12.2 & 13.02 & 0.005 & H II  & & NGC 3504          \\
 11 &  8 & 43.37 & 29 & 13 & 25.3 &   2.7 & 0.022 &42.76& 13.4 & 16.49 & 0.078 & em    & 1,2 & LEDA 094150       \\
 11 & 10 & 45.96 & 11 & 36 & 41.2 &  14.3 & 0.025 &42.28& 13.2 & 16.16 & 0.042 & S1    & 1 & Zw 067.027        \\
\hline
\end{tabular}
\end{center}
\end{table*}
\addtocounter{table}{-1}
\begin{table*}
\caption{(continued)}
\begin{center}
\begin{tabular}{rrr|rrr|r|cc|cc|c|l|l|l}
\hline \multicolumn{6}{c}{J2000 optical position} \\
\hline
h & m  & s & $^{\circ}$ & ' & " & sep(") & c/r & L$_{\rm X}$ & O$_{\rm US}$ & O$_{\rm APS}$& z &  & ref. & Name  \\
\hline 
 11 & 15 &  2.23 &  4 &  5 &  7.4 &  13.4 & 0.024 &42.19& 13.4 & 16.48 & 0.039 &       & & Zw 039.090        \\
 11 & 24 & 43.64 & 38 & 45 & 46.3 &  22.1 & 0.051 &41.03& 12.9 & 12.50 & 0.007 & H II  & & NGC 3665          \\
 11 & 25 & 45.33 & 24 &  8 & 24.5 &  28.9 & 0.029 &41.86& 12.7 & 15.12 & 0.024 &       & & Zw 126.051        \\
 11 & 28 &  2.16 & 78 & 59 & 40.4 &  16.3 & 0.046 &     & 14.7 &       & 0.000 & H II  & & UGC  6456         \\
 11 & 34 &  6.64 & 25 & 33 & 34.8 &  18.9 & 0.029 &     & 12.8 & 16.55 &       &       & & MCG+04.27.068     \\
 11 & 39 & 14.88 & 17 &  8 & 37.2 &   4.2 & 0.037 &41.85& 13.1 & 16.23 & 0.021 &       & & UGC  6614         \\
 11 & 39 & 41.57 & 31 & 54 & 41.1 &   2.9 & 0.034 &41.07& 15.9 &       & 0.009 & S1.8  & & NGC 3786          \\
 11 & 46 & 12.24 & 20 & 23 & 29.9 &  27.2 & 0.048 &42.04& 14.5 & 14.61 & 0.023 & Liner & & NGC 3884          \\
 11 & 47 & 22.16 & 35 &  1 &  7.5 &  25.9 & 0.044 &42.87& 14.4 & 16.60 & 0.063 & S2    & & B2 1144+35B       \\
 11 & 49 & 59.31 & 21 & 20 &  1.1 &   9.2 & 0.026 &41.88&      & 15.32 & 0.026 &       & & NGC 3910          \\
 11 & 50 & 41.53 & 20 &  0 & 54.6 &  10.2 & 0.031 &41.77&      & 15.55 & 0.021 &       & & NGC 3919          \\
 11 & 53 & 26.26 & 35 & 56 & 53.8 &  15.3 & 0.023 &42.66& 12.3 & 16.27 & 0.068 & S2    & 2 & KUG 1150+362    \\
 11 & 59 & 52.16 & 55 & 32 &  5.4 &  30.6 & 0.024 &42.83& 13.7 & 16.96 & 0.081 &       & & PGC 2505536       \\
 12 &  4 &  6.25 & 20 & 14 &  6.2 &  25.5 & 0.028 &41.72& 12.9 & 14.99 & 0.021 &       & & NGC 4065          \\
 12 &  4 & 43.34 & 31 & 10 & 38.2 &  16.7 & 0.025 &41.83&      & 15.11 & 0.025 & S1.9  & & UGC  7064         \\
 12 &  5 & 49.83 & 35 & 10 & 46.4 &  11.9 & 0.190 &43.38& 12.7 & 16.74 & 0.054 & S1    & & Mark 646          \\
 12 &  8 &  5.56 & 25 & 14 & 14.1 &  16.0 & 0.034 &41.89&      & 15.32 & 0.023 &       & & UGC  7115         \\
 12 & 10 & 33.61 & 30 & 24 &  5.9 &  21.4 & 0.045 &39.29&      & 13.71 & 0.001 & C     & & NGC 4150          \\
 12 & 14 & 18.10 & 29 & 31 &  4.5 &   5.0 & 0.021 &42.57&      & 16.96 & 0.064 & S2    & & WAS 49b           \\
 12 & 14 & 48.65 & 59 & 54 & 22.5 &  23.3 & 0.028 &42.64& 14.6 & 15.61 & 0.060 &       & & NGC 4199a         \\
 12 & 14 & 51.71 & 59 & 54 & 30.5 &   4.5 & 0.028 &42.65& 16.3 &       & 0.061 & abs   & 1 & NGC 4199b         \\
 12 & 15 &  5.34 & 76 & 14 & 10.0 &  34.4 & 0.031 &40.68&      &       & 0.006 &       & & UGC 7265          \\
 12 & 16 &  0.00 & 12 & 41 &  1.4 &   5.4 & 0.033 &42.78&      & 16.60 & 0.065 & S1.9  & & Mark  764         \\
 12 & 19 & 51.66 & 28 & 25 & 21.7 &  12.6 & 0.046 &42.13& 13.8 & 16.27 & 0.026 &       & & Zw 158.075        \\
 12 & 20 &  6.86 & 29 & 16 & 50.5 &  11.4 & 0.037 &39.80& 12.6 & 12.53 & 0.002 & Liner & & NGC 4278          \\
 12 & 22 & 54.89 & 15 & 49 & 20.7 &  15.6 & 0.042 &40.65&      & 11.30 & 0.005 & C     & & NGC 4321          \\
 12 & 23 & 39.10 &  7 &  3 & 14.0 &  24.6 & 0.030 &40.06&      & 14.80 & 0.003 &       & & NGC 4342          \\
 12 & 23 & 42.85 & 58 & 14 & 45.9 &  10.3 & 0.037 &41.55& 14.5 & 16.93 & 0.015 & S?    & & SBS 1221+585      \\
 12 & 26 & 28.02 &  9 &  1 & 23.0 &  26.7 & 0.029 &41.86& 14.7 & 14.58 & 0.024 & Liner & & NGC 4410          \\
 12 & 26 & 26.98 & 31 & 13 & 22.6 &   4.4 & 0.033 &39.75& 12.2 & 12.17 & 0.002 & C     & & NGC 4414          \\
 12 & 28 & 29.49 & 17 &  5 &  6.0 &   9.8 & 0.039 &40.91&      & 12.21 & 0.007 & Liner & & NGC 4450          \\
 12 & 34 &  3.10 &  7 & 41 & 59.0 &  24.2 & 0.032 &39.14&      & 11.85 & 0.001 & H II  & & NGC 4526          \\
 12 & 34 & 46.81 & 47 & 45 & 32.4 &  34.2 & 0.043 &42.25&      & 16.12 & 0.031 &       & & MCG+08.23.061     \\
 12 & 48 & 58.34 & 40 & 35 & 56.8 &  14.5 & 0.039 &41.58& 15.2 & 15.95 & 0.015 &       & & IC 3808           \\
 12 & 50 & 26.60 & 25 & 30 &  5.8 &   8.3 & 0.028 &40.28&      & 11.04 & 0.004 & S2    & & NGC 4725          \\
 12 & 55 &  7.78 & 78 & 37 & 14.9 &   4.1 & 0.029 &42.36& 12.9 &       & 0.043 & S1    & 1 & NPM1G+78.0053     \\
 12 & 56 & 43.76 & 21 & 40 & 51.9 &   7.2 & 0.041 &39.25&      & 10.36 & 0.001 & S     & & NGC 4826          \\
 12 & 59 & 39.35 & 38 & 48 & 56.3 &  16.0 & 0.022 &42.01& 13.2 & 16.23 & 0.033 &       & & IC 4003           \\
 13 &  0 & 39.13 &  2 & 30 &  5.3 &  22.4 & 0.027 &40.02&      & 12.98 & 0.003 & H II  & & NGC 4900          \\
 13 &  4 & 40.89 & 43 & 18 & 35.0 &  20.1 & 0.031 &     & 14.0 & 16.86 &       &       & & NPM1G+43.0235     \\
 13 &  4 & 57.99 & 43 & 33 & 10.9 &  25.2 & 0.052 &     & 13.1 & 16.06 &       &       & & MCG+07.27.026     \\
 13 &  7 &  3.03 & 56 & 31 & 59.2 &   5.5 & 0.044 &43.08& 12.4 & 16.40 & 0.080 & S1    & 1,2 & MCG+10.19.033     \\
 13 & 20 & 28.94 & 31 & 21 &  6.9 &  19.8 & 0.029 &42.42& 12.8 & 16.53 & 0.046 &       & & Zw 160.204        \\
 13 & 29 & 58.73 & 47 & 16 &  4.5 &  29.1 & 0.046 &39.90&      & 11.51 & 0.002 & Liner & & NGC 5195          \\
 13 & 32 & 48.62 & 41 & 52 & 18.9 &  12.3 & 0.038 &42.07&      & 14.08 & 0.027 &       & & NGC 5214          \\
 13 & 38 & 17.27 & 48 & 16 & 32.0 &  18.4 & 0.043 &42.16&      & 15.05 & 0.028 & S2    & & Mark 266SW        \\
 13 & 42 &  8.34 & 35 & 39 & 15.2 &  15.3 & 0.021 &40.16&      & 13.06 & 0.004 & S1.9  & & NGC 5273          \\
 13 & 44 &  1.90 & 25 & 56 & 27.8 &  28.5 & 0.026 &     & 12.8 & 17.00 &       &       & &                   \\
 13 & 51 &  4.24 & 19 & 26 &  8.2 &   3.7 & 0.048 &42.90& 13.5 & 16.95 & 0.062 & em    & 1,2 &                   \\
 13 & 51 & 42.19 & 55 & 59 & 43.2 &  29.3 & 0.022 &     & 15.1 & 16.68 &       &       & & TEX 1349+562      \\
 13 & 53 &  9.75 & 24 & 22 & 37.4 &   6.6 & 0.027 &42.26& 13.5 & 16.85 & 0.056 & S1    & 1,2 & Zw 132.035        \\
 13 & 53 & 26.69 & 40 & 16 & 58.9 &   7.7 & 0.048 &41.12&      &       & 0.008 & Liner & & NGC 5353          \\
 13 & 53 & 43.82 & 33 & 13 & 20.3 &  28.2 & 0.024 &42.43& 14.5 & 15.61 & 0.051 &       & & MCG +06.31.015    \\
 13 & 53 & 55.94 & 21 & 59 & 54.9 &  24.1 & 0.042 &     & 12.9 & 16.86 &       &       & &                   \\
 13 & 58 & 19.76 &  7 & 13 & 15.2 &  27.4 & 0.044 &42.07& 15.7 &       & 0.025 &       & & Zw 046.025        \\
 14 & 12 & 36.95 & 39 & 18 & 53.8 &  21.2 & 0.034 &41.96& 16.4 & 15.27 & 0.025 & S1.9  & & NGC 5515          \\
 14 & 13 & 23.13 & 24 & 31 & 56.8 &  24.3 & 0.024 &42.44& 12.2 & 16.34 & 0.052 & abs   & 2 & Zw 133.012        \\
 14 & 18 &  9.20 &  7 & 33 & 52.0 &  24.3 & 0.042 &42.02&      & 14.97 & 0.024 &       & & NGC 5546          \\
\hline
\end{tabular}
\end{center}
\end{table*}
\addtocounter{table}{-1}
\begin{table*}
\caption{(continued)}
\begin{center}
\begin{tabular}{rrr|rrr|r|cc|cc|c|l|l|l}
\hline \multicolumn{6}{c}{J2000 optical position} \\
\hline
h & m  & s & $^{\circ}$ & ' & " & sep(") & c/r & L$_{\rm X}$ & O$_{\rm US}$ & O$_{\rm APS}$ & z & & ref. & Name  \\
\hline
 14 & 22 & 55.37 & 32 & 51 &  2.7 &   2.0 & 0.028 &42.14&      & 15.28 & 0.034 & S1.8  & & UGC  9214         \\
 14 & 26 &  2.32 &  8 &  6 & 40.5 &  34.6 & 0.024 &41.84&      & 16.98 & 0.026 &       & & Zw 047.039        \\
 14 & 26 & 18.64 & 26 & 14 & 52.3 &  13.8 & 0.027 &42.02& 16.3 & 15.58 & 0.030 &       & & IC 4423           \\
 14 & 28 & 23.06 & 78 & 53 &  8.2 &   0.2 & 0.047 &42.03& 12.1 &       & 0.023 &       & & IC 4470           \\  
 14 & 29 & 40.64 &  0 & 21 & 59.2 &  13.8 & 0.025 &42.51& 14.8 & 16.92 & 0.055 & abs   & 1 & Zw 019.045      \\
 14 & 29 & 48.51 & 53 & 57 & 54.3 &  25.6 & 0.025 &42.30& 13.2 & 16.19 & 0.043 &       & & IC 1027           \\
 14 & 42 & 39.65 & 29 & 20 & 48.2 &   7.5 & 0.028 &42.82& 13.5 & 17.00 & 0.074 & H II  & 1,2 & NPM1G+29.0326     \\
 14 & 50 & 51.40 &  5 &  6 & 52.0 &  25.9 & 0.025 &41.92&      & 15.29 & 0.028 & S2    & & NGC 5765B         \\
 14 & 50 & 55.42 & 27 & 34 & 42.1 &  13.1 & 0.032 &42.09&      & 15.89 & 0.030 &       & & IC 4514           \\
 14 & 51 & 14.39 & 30 & 41 & 32.2 &  33.9 & 0.044 &     & 12.5 & 16.28 &       &       & & Zw 164.036        \\
 14 & 55 & 28.21 & 32 & 50 & 24.0 &  30.3 & 0.026 &     & 14.1 & 16.21 &       &       & & MCG+06.33.009     \\
 15 &  4 & 15.88 & 28 & 29 & 47.5 &  32.7 & 0.127 &43.26&      & 15.66 & 0.058 &       & & MCG+05.36.002     \\
 15 &  5 & 56.60 &  3 & 42 & 26.0 &  15.6 & 0.027 &42.18&      & 16.00 & 0.036 & S1.8  & & Mark 1392         \\
 15 &  9 & 20.51 &  7 & 38 & 19.4 &  28.5 & 0.044 &43.05& 16.0 & 16.94 & 0.077 & abs   & 2 &                   \\
 15 & 13 & 40.25 &  4 &  4 & 17.2 &  24.9 & 0.025 &42.21& 13.3 & 16.96 & 0.039 & H II  & 2 & Zw 049.061        \\
 15 & 20 & 29.00 & 44 & 58 & 15.3 &  11.5 & 0.046 &42.89& 13.9 & 16.86 & 0.063 & abs   & 1 &                   \\
 15 & 21 & 20.58 & 30 & 40 & 15.3 &  15.0 & 0.139 &43.57& 14.0 & 16.93 & 0.079 & abs   & 2 & Zw 165.041        \\
 15 & 23 & 59.90 & 31 & 12 & 40.2 &  31.4 & 0.032 &42.97&      & 16.71 & 0.074 &       & & MCG+05.36.026     \\
 15 & 25 &  8.75 & 12 & 52 & 57.5 &  24.6 & 0.039 &41.95& 12.5 & 16.30 & 0.023 &       & & Zw 077.123        \\
 15 & 29 & 14.68 & 52 & 51 & 50.1 &  12.9 & 0.029 &42.80& 13.1 & 16.66 & 0.071 & abs   & 1 & UGC  9868         \\
 15 & 32 & 32.02 &  4 & 40 & 51.4 &  13.8 & 0.048 &42.50&      & 15.75 & 0.039 &       & & UGC  9886         \\
 15 & 32 & 57.19 &  0 & 26 & 36.1 &  11.4 & 0.044 &42.41& 15.8 & 16.11 & 0.037 &       & & Zw 022.010        \\
 15 & 33 & 17.73 & 82 & 13 & 46.2 &  11.3 & 0.036 &41.87& 12.9 &       & 0.022 &       & & UGC 9950          \\
 15 & 35 & 54.26 & 14 & 31 &  2.7 &  34.6 & 0.049 &41.92&      & 15.70 & 0.020 & S2    & & Akn 479           \\
 15 & 38 & 10.00 & 57 & 36 & 12.0 &  12.1 & 0.034 &42.90&      & 16.20 & 0.074 & S1    & & MCG+10.22.028     \\
 15 & 39 &  2.65 & 31 & 45 & 34.3 &  25.4 & 0.031 &40.81&      & 15.31 & 0.007 &       & & NGC 5974          \\
 15 & 47 & 48.93 & 37 &  1 & 36.0 &  19.9 & 0.040 &43.01& 15.1 & 16.60 & 0.077 & em    & 2 & NPM1G+37.0489     \\
 15 & 52 & 12.04 & 34 &  5 & 35.4 &  24.0 & 0.024 &42.43& 14.7 &       & 0.051 & abs   & 2 & NPM1G+34.0353     \\
 15 & 56 & 41.39 & 20 & 10 & 16.4 &  33.0 & 0.027 &     & 12.8 & 16.66 &       &       & &                   \\
 16 &  4 & 55.53 & 28 &  9 & 56.9 &  21.8 & 0.047 &43.08& 12.1 & 16.54 & 0.077 &       & & Zw 167.022        \\
 16 &  5 & 29.18 & 16 & 25 &  8.8 &  28.3 & 0.041 &42.53&      &       & 0.044 &       & & UGC 10187B        \\
 16 &  5 & 35.57 & 44 & 12 & 22.0 &  16.9 & 0.056 &42.65& 13.5 & 16.33 & 0.043 &       & & MCG+07.33.038     \\
 16 &  5 & 36.83 & 17 & 48 &  8.1 &  32.0 & 0.025 &42.09&      & 16.32 & 0.034 & H II  & & Mark  298         \\
 16 &  7 & 24.03 & 85 &  1 & 49.3 &   2.7 & 0.042 &43.78& 16.5 &       & 0.183 & S1    & & S5 1616+85        \\
 16 &  7 & 35.36 & 13 & 56 & 37.6 &  28.9 & 0.026 &42.11& 13.1 & 16.21 & 0.034 &       & & NGC 6066          \\
 16 &  9 &  5.53 & 27 & 53 & 34.0 &  10.2 & 0.041 &42.23& 12.5 & 16.84 & 0.031 & S2    & 1,2 & NPM1G+28.0373     \\
 16 &  9 & 35.18 & 63 & 58 &  1.2 &   8.2 & 0.047 &42.75& 14.0 & 16.82 & 0.053 & em    & 1,2 & NPM1G+64.0143     \\
 16 & 11 & 11.40 & 61 & 16 &  4.5 &  13.2 & 0.030 &42.09&      & 14.95 & 0.031 &       & & NGC 6095          \\
 16 & 11 & 13.89 & 36 & 58 & 24.3 &  27.6 & 0.031 &42.79& 12.5 & 16.95 & 0.068 & abs   & 2 & KUG 1609+371A     \\
 16 & 12 & 33.72 & 29 & 29 & 39.1 &  15.5 & 0.043 &42.28& 13.6 & 15.07 & 0.032 &       & & NGC 6086          \\
 16 & 24 & 37.00 & 19 & 30 & 24.3 &  11.2 & 0.026 &42.16&      &       & 0.036 &       & & Zw 109.013        \\
 16 & 27 & 42.60 & 39 & 22 & 38.5 &  13.7 & 0.023 &41.92& 16.1 &       & 0.029 &       & & PGC 058195        \\
 16 & 29 & 44.90 & 40 & 48 & 41.8 &   6.2 & 0.039 &42.15&      & 14.31 & 0.029 &       & & NGC 6173          \\
 16 & 29 & 52.80 & 24 & 26 & 38.1 &   6.3 & 0.041 &42.40& 12.2 & 16.30 & 0.038 & S1.9  & & Mark  883         \\
 16 & 32 & 58.03 & 11 & 43 & 23.8 &  27.5 & 0.038 &42.71& 16.3 & 15.65 & 0.056 &       & & Zw 080.046        \\
 16 & 33 & 48.87 & 35 & 53 & 18.8 &  33.3 & 0.022 &42.13&      & 16.57 & 0.038 &       & & KUG 1632+359      \\
 16 & 36 & 57.72 & 55 & 46 & 58.2 &   8.6 & 0.038 &42.11& 13.5 & 16.92 & 0.028 & S?    & 1,2 &                   \\
 16 & 37 & 20.55 & 41 & 11 & 20.1 &  16.3 & 0.038 &42.22& 13.0 & 16.74 & 0.032 & abs   & 2 & NPM1G+41.0441     \\
 16 & 39 &  4.72 &  8 & 21 & 30.9 &  11.2 & 0.035 &42.48& 13.2 & 16.88 & 0.045 & em    & 2 & NPM1G+08.0448     \\
 16 & 42 & 56.33 & 19 & 15 & 15.5 &  10.6 & 0.042 &42.32& 13.0 & 16.25 & 0.034 &       & & IC 1224           \\
 16 & 43 &  4.22 & 61 & 34 & 43.3 &  22.7 & 0.032 &41.70& 13.9 & 14.13 & 0.019 &       & & NGC 6223          \\
 16 & 56 &  1.58 & 21 & 12 & 42.0 &  10.4 & 0.028 &42.46& 15.4 & 16.99 & 0.049 & S1    & 1 & NPM1G+21.0507     \\
 16 & 57 & 45.01 & 68 & 30 & 53.1 &  13.9 & 0.045 &42.42& 13.1 & 15.88 & 0.037 &       & & NGC 6289          \\
 17 &  0 & 27.21 & 51 & 59 & 11.6 &  14.3 & 0.044 &42.64& 12.7 & 16.87 & 0.048 & S1    & & NPM1G+52.0273     \\
 17 &  3 & 47.89 & 34 & 43 & 39.2 &  20.0 & 0.023 &     & 12.2 &       &       &       & &                   \\
 17 & 12 & 36.58 & 38 &  1 & 13.3 &  25.4 & 0.040 &42.37&      & 15.40 & 0.037 &       & & IC 1245           \\
 17 & 15 & 58.85 & 36 & 23 & 23.1 &  15.2 & 0.048 &43.18& 13.6 & 16.13 & 0.086 & S1    & 2 & UGC 10782         \\
 17 & 23 & 22.08 & 32 & 49 & 55.2 &  16.7 & 0.022 &42.63& 13.2 & 16.55 & 0.067 & H II  & 2 &                   \\
\hline
\end{tabular}
\end{center}
\end{table*}
\addtocounter{table}{-1}
\begin{table*}
\caption{(end)}
\begin{center}
\begin{tabular}{rrr|rrr|r|cc|cc|c|l|l|l}
\hline \multicolumn{6}{c}{J2000 optical position} \\
\hline
h & m  & s & $^{\circ}$ & ' & " & sep(") & c/r & L$_{\rm X}$ & O$_{\rm US}$ & O$_{\rm APS}$ & z & & ref. & Name  \\
\hline  
 17 & 37 & 56.34 & 41 & 38 & 32.0 &  19.9 & 0.023 &42.68& 15.1 & 16.39 & 0.070 & H II  & 2 & IRAS F17363+4140  \\
 17 & 38 & 11.38 & 58 & 42 & 55.3 &  21.7 & 0.034 &42.09&      & 15.85 & 0.029 & S1    & & NGC 6418          \\
 22 & 21 & 47.39 &  2 & 54 & 36.2 &  24.5 & 0.028 &     & 13.2 & 16.72 &       &       & &                   \\
 22 & 28 & 29.51 & 16 & 46 & 59.6 &  11.6 & 0.034 &42.23& 12.9 & 15.01 & 0.034 &       & & NGC 7291          \\
 22 & 32 & 30.68 &  8 & 12 & 33.0 &  18.2 & 0.034 &41.96& 12.5 & 16.08 & 0.025 & S1   & & Akn 557           \\
 22 & 35 & 40.82 &  1 & 29 &  5.9 &  26.2 & 0.293 &43.64& 14.3 & 16.63 & 0.059 &       & & LEDA 087323       \\
 22 & 41 & 34.22 &  4 & 53 & 10.7 &  34.3 & 0.047 &42.97&      & 16.90 & 0.068 & S?    & & NPM1G+04.0574     \\
 22 & 49 & 54.69 & 11 & 36 & 30.1 &  23.7 & 0.054 &42.19&      & 14.35 & 0.026 &       & & NGC 7385          \\
 22 & 58 &  1.97 & 13 &  8 &  4.4 &  10.7 & 0.021 &41.75&      & 15.08 & 0.025 &       & & NGC 7432          \\
 23 & 13 & 58.36 &  3 & 42 & 54.4 &   5.6 & 0.031 &42.62& 13.0 & 16.42 & 0.056 &       & & Zw 380.012        \\
 23 & 20 & 42.29 &  8 & 13 &  2.5 &   7.2 & 0.035 &41.26&      &       & 0.011 & Liner & & NGC 7626          \\
 23 & 25 & 51.50 &  8 & 47 & 11.2 &   6.6 & 0.025 &     & 14.3 & 16.46 &       &       & & KUG 2323+085      \\
 23 & 31 & 50.20 & 25 & 32 & 40.0 &   6.7 & 0.038 &     & 13.3 & 16.59 &       &       & & KUG 2329+252      \\
 23 & 36 & 14.10 &  2 &  9 & 18.6 &   3.5 & 0.021 &40.86&      &       & 0.009 & H II  & & NGC 7714          \\
 23 & 56 &  1.96 &  7 & 31 & 23.4 &   3.1 & 0.049 &42.53& 13.6 & 16.84 & 0.040 & S1    & & Mark 541          \\
\hline
\end{tabular}
\end{center}
\end{table*}
\normalsize

\begin{figure}[ht]

\resizebox{8.8cm}{!}{\includegraphics{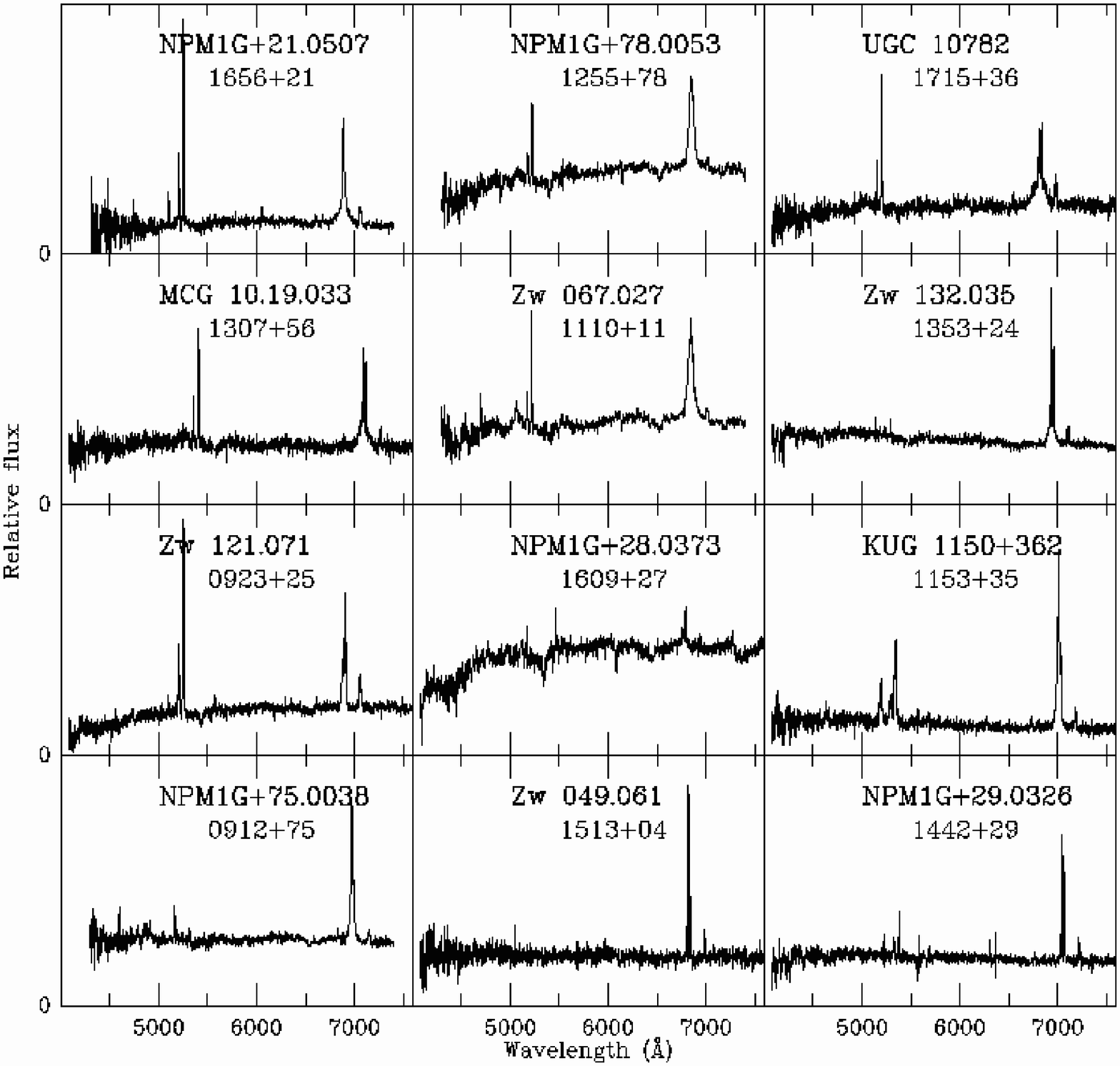}}

\caption{\label{spec_gal} Spectra of the six Seyfert 1, the three Seyfert 2 and 
three of the five starburst galaxies confirmed by our spectroscopic observations 
at BAO and OHP.}

\end{figure}

 We have observed 34 of the galaxies without a previously known redshift at the 
BAO and OHP observatories. The BAO observations were made during the period 
March 3 to 10, 2002 with the 2.6-m telescope using the SCORPIO spectral camera 
attached to the prime focus. The combination of the detector, a 
2063$\times$2058 16$\times$16 $\mu$m pixel Loral Lick3 CCD, with a 600 g mm$^{-1}$
grism resulted in the spectral range 3900-7400 \AA\ with a dispersion of 1.7 \AA\
pixel$^{-1}$. The slit width was 1$\farcs$8 (4.3 pixels) and was oriented EW.
The spectral resolution was 5 \AA\ FWHM. The exposure time was usually 45 min. 
The OHP observations were made 
between April 15 and 19, 2002 with the 1.93-m telescope and the spectrograph 
CARELEC (Lema\^{\i}tre et al. \cite{lemaitre89}) attached to the Cassegrain 
focus. The detector was a 2048$\times$1024 13.5$\times$13.5 $\mu$m pixel
EEV CCD. The dispersion was 1.75 \AA\ pixel$^{-1}$ and the spectral range 4075-7715
\AA. The slit width was 2$\farcs$0, corresponding to 3.7 pixels, and the 
resolution was 5.7 \AA\ FWHM. The exposure time was usually 20 min. We found 13 
absorption line galaxies, five starburst 
galaxies, six Seyfert 1, three Seyfert 2 and seven unclassifiable weak emission 
line galaxies. Fig. \ref{spec_gal} shows the spectra of the Seyfert 1 and 2 
galaxies and of three starburst galaxies.

 In total the redshift of 164 of the galaxies is known. We have computed their
0.1-2.4 keV X-ray luminosity. Sixty-three are smaller than 10$^{42}$ erg s$^{-1}$
and forty-five, greater than 10$^{42.5}$ erg s$^{-1}$.  
As there are probably no starburst galaxies having an 0.1-2.4 keV X-ray luminosity 
substantially higher than 10$^{42}$ erg s$^{-1}$ (Moran et al. \cite{moran96}; 
Condon et al. \cite{condon98}), most of the 45 high X-ray luminosity 
galaxies must be AGN. Among the 16 which have been classified, four are 
starbursts and 12 AGN. 
The 63 low X-ray luminosity galaxies could possibly be absorbed 
Seyfert 2s, low-luminosity AGN as well as starbursts, or elliptical 
galaxies with hot gas (Hornschemeier et al. \cite{hornschemeier01}; 
\cite{hornschemeier03}; Comastri et al. \cite{comastri02}; Severgnini et al. 
\cite{severgnini03}). Several of our low X-ray luminosity sources have 
been classified as Liners, Seyfert 2s, Seyfert 1.9 or 1.8.

\section{QSOs}

\begin{figure}[ht]

\resizebox{8.8cm}{!}{\includegraphics{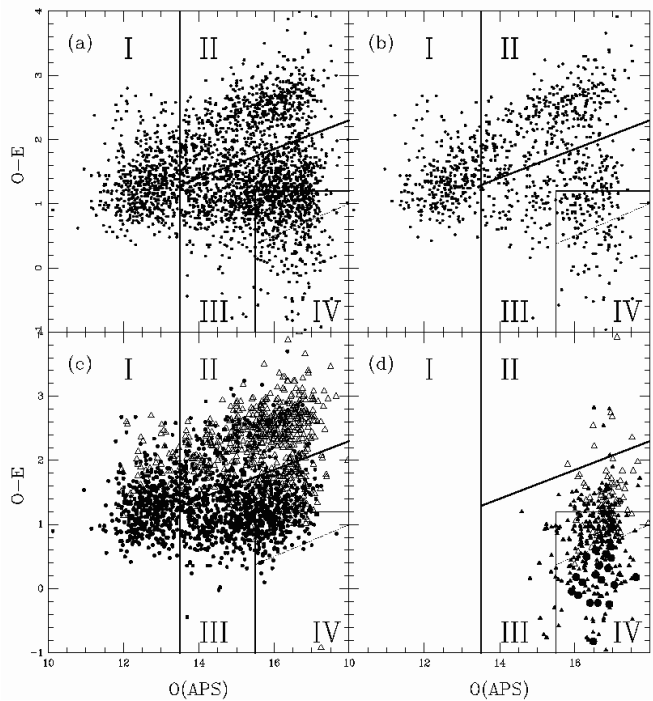}}

\caption{\label{colmag} (a) APS O--E colour indices $\it vs$ O magnitudes for 
all starlike candidate identifications within 35$\arcsec$; (b) the same for 
coincidences within 15$\arcsec$; (c) the same for the A to G stars (filled circles)
and the K and M stars (open triangles) as classified according to the Hamburg scheme; 
(d) the same for the BLUE-WK and EBL-WK objects (filled triangles), the RED-WK objects 
(open triangles) and the QSOs (filled circles).}

\end{figure}

 Figure \ref{colmag} (a) shows the APS O--E colour index $\it vs$ the O magnitude 
for the 2\,652 starlike objects in the 3\,212 sample. Fig. \ref{colmag} (b) is the 
same for objects within 15$\arcsec$ of a ROSAT-FSC source. These figures were 
divided into four zones.

\begin{table*}
\caption{\label{xrayq} List of the 76 known AGN brighter 
than O$_{\rm USNO}$=16.5, located within 35$\arcsec$ of a ROSAT-FSC source. Col. 1: 
right ascension, col. 2: declination, col. 3: separation between 
the ROSAT and USNO positions in arcsec, col. 4: 
X-ray count rate (count s$^{-1}$), col. 5: magnitude O$_{\rm USNO}$, col. 6: magnitude 
O$_{\rm APS}$, col. 7: APS O--E colour, col. 8: redshift, col. 9: absolute magnitude 
computed by using the O$_{\rm USNO}$ magnitude (H$_{\rm o}$=50 km s$^{-1}$ Mpc$^{-1}$), 
col. 10: classification, col. 11: name.}
\begin{center}
\begin{tabular}{rrr|rrr|r|c|rrr|l|r|l|l}
\hline \multicolumn{6}{c}{J2000 USNO position} \\
\hline
h & m  & s & $^{\circ}$ & ' & " & " & c/r & O$_{\rm US}$ & O$_{\rm APS}$ & O--E & z & M$_{\rm abs}$ & &  Name  \\
\hline

  0 &  6 & 23.08 & 12 & 35 & 53.2 &  20.8 & 0.030 & 15.8 & 16.26 &   0.36 & 0.98  & --28.4 & Q      & RGB J0006+125        \\
  0 & 24 & 44.10 &  0 & 32 & 21.4 &   9.7 & 0.047 & 16.2 & 16.92 &   0.58 & 0.404 & --25.8 & Q      & PB 5932              \\
  0 & 36 & 12.50 &  5 & 49 & 52.0 &   9.1 & 0.064 & 16.3 & 16.88 &   0.28 & 0.41  & --25.7 & Q      & HS 0033+0533         \\
  0 & 52 &  5.57 &  0 & 35 & 38.4 &  28.6 & 0.024 & 16.3 & 16.80 &   1.04 & 0.399 & --25.7 & Q      & Q 0049+0019A         \\
  0 & 57 &  9.93 & 14 & 46 & 10.4 &  26.7 & 0.246 & 15.9 & 16.11 &   1.12 & 0.171 & --24.0 & Q      & PHL 909              \\
  1 & 20 & 17.29 & 21 & 33 & 46.6 &  25.0 & 0.023 & 15.5 & 15.80 &   0.54 & 1.500 & --29.7 & Q      & PG 0117+213          \\
  1 & 39 & 55.80 &  6 & 19 & 22.5 &   8.4 & 0.031 & 15.7 & 16.55 &   0.60 & 0.396 & --26.5 & Q      & PHL 1092             \\
  1 & 40 & 35.02 & 23 & 44 & 51.1 &  25.1 & 0.029 & 16.1 & 16.64 &   0.98 & 0.32  & --25.4 & Q      & HS 0137+2329         \\
  1 & 41 & 59.60 &  6 & 12 &  5.5 &   2.7 & 0.049 & 16.3 & 16.27 &   0.56 & 0.345 & --27.2 & Q      & PHL 1106             \\
  3 & 18 & 25.57 & 15 & 59 & 56.7 &  11.2 & 0.033 & 16.3 & 16.58 &   0.02 & 0.515 & --25.9 & Q      & US 3828              \\
  3 & 39 &  9.60 &  3 & 45 & 52.5 &  23.7 & 0.038 & 16.4 & 17.55 &   0.44 & 0.199 & --24.0 & Q      & RXS J03391+0346      \\
  4 & 24 & 46.85 &  0 & 36 &  6.4 &   7.4 & 0.052 & 16.1 & 16.75 &   1.26 &       &        & BL     & PKS 0422+00          \\
  8 & 15 & 59.21 & 41 & 44 & 57.1 &  17.9 & 0.021 & 15.8 & 15.97 &   0.32 & 1.28  & --29.3 & Q      & KUV 08126+4154       \\
  8 & 22 & 36.88 & 54 & 18 & 36.4 &  33.7 & 0.042 & 16.3 & 16.65 &   1.12 & 0.086 & --22.3 & S1     & SBS 0818+544         \\
  8 & 43 & 49.78 & 26 & 19 & 11.1 &  19.4 & 0.068 & 16.4 & 17.23 &   0.20 & 0.258 & --24.7 & Q      & HS 0840+2630         \\
  8 & 59 & 24.35 & 46 & 37 & 17.3 &  22.3 & 0.047 & 16.5 & 16.89 &   0.74 & 0.923 & --27.5 & Q      & US 2068              \\
  9 &  0 & 47.30 & 74 & 44 & 26.3 &  16.6 & 0.045 & 16.1 &       &        & 0.77  & --27.4 & Q      & HS 0855+7456         \\
  9 & 19 & 57.63 & 51 &  6 &  9.1 &  17.1 & 0.039 & 15.8 & 16.35 & --0.80 & 0.553 & --27.0 & Q      & SBS 0916+513         \\
  9 & 25 & 14.35 & 54 & 44 & 27.2 &  10.4 & 0.048 & 15.9 & 16.66 & --0.12 & 0.476 & --26.3 & Q      & SBS 0921+549         \\
  9 & 29 &  9.82 & 46 & 44 & 24.1 &  15.9 & 0.025 & 15.6 & 15.66 &   0.18 & 0.240 & --25.3 & Q      & US 645               \\
  9 & 36 & 25.45 & 39 & 49 & 33.7 &  24.8 & 0.033 & 16.1 & 16.81 &   0.82 & 1.25  & --28.7 & Q      & KUV 09333+4003       \\
  9 & 37 &  1.93 & 34 & 25 &  0.0 &  12.7 & 0.031 & 16.0 & 16.55 &   0.22 & 0.908 & --28.1 & Q      & Ton 1078             \\
  9 & 41 & 33.72 & 59 & 48 & 11.3 &  13.2 & 0.045 & 16.3 & 16.70 &   0.36 & 0.966 & --27.9 & Q      & SBS 0938+600         \\
  9 & 56 & 49.89 & 25 & 15 & 16.0 &   8.0 & 0.048 & 16.2 & 16.92 &   0.84 & 0.712 & --27.2 & Q      & OK 290               \\
 10 &  4 & 20.10 &  5 & 13 &  0.6 &  22.1 & 0.021 & 16.2 & 16.55 &   0.26 & 0.161 & --23.8 & Q      & PG 1001+054          \\
 10 & 13 & 30.20 & 53 & 15 & 59.6 &   6.7 & 0.025 & 16.2 & 16.64 &   0.36 & 1.495 & --29.1 & Q      & SBS 1010+535         \\
 10 & 15 & 57.04 &  1 &  9 & 13.7 &  16.1 & 0.082 & 16.3 & 16.97 &   0.94 & 0.779 & --27.2 & Q      & Q 1013+0124          \\
 10 & 33 & 59.50 & 35 & 55 &  9.0 &  14.9 & 0.046 & 16.3 & 16.76 &   0.92 & 0.169 & --23.8 & Q      & CSO 275              \\
 10 & 43 & 55.52 & 56 & 27 & 57.0 &  18.6 & 0.028 & 16.3 & 16.81 & --0.40 & 1.951 & --29.6 & Q      & SBS 1040+567         \\
 11 & 17 &  6.41 & 44 & 13 & 33.8 &   8.1 & 0.023 & 14.8 & 15.06 &   0.40 & 0.144 & --24.9 & Q      & PG 1114+445          \\
 11 & 30 &  4.76 & 41 & 16 & 19.5 &  23.2 & 0.035 & 15.7 & 16.20 &   0.10 & 1.530 & --29.5 & Q      & KUV 11274+4133       \\
 11 & 33 & 35.40 &  9 & 39 &  1.8 &  24.6 & 0.029 & 16.4 & 17.05 &   0.06 & 0.379 & --25.5 & Q      & RX J11335+0939       \\
 11 & 43 & 47.71 & 11 & 28 & 48.1 &  20.6 & 0.046 & 15.5 & 16.38 &   0.80 & 0.118 & --23.8 & Q      & RX J11437+1128       \\
 11 & 52 & 51.90 & 33 &  7 & 18.8 &   2.7 & 0.033 & 16.1 & 16.30 &   0.40 & 1.389 & --28.8 & Q      & CSO 373              \\
 11 & 55 &  7.62 & 52 &  1 & 29.4 &   6.4 & 0.036 & 16.4 & 16.46 &   1.08 & 0.156 & --23.5 & Q      & SBS 1152+523         \\
 11 & 59 &  6.78 & 53 &  6 & 43.5 &   6.1 & 0.043 & 16.5 & 17.19 &   0.36 & 0.482 & --25.9 & Q      & MS 11565+5323        \\
 12 &  7 &  4.53 & 38 & 40 & 24.6 &   3.7 & 0.045 & 16.3 & 16.64 &   0.00 & 0.572 & --26.4 & Q      & RXS J12070+3840      \\
 12 & 17 & 40.83 & 49 & 31 & 17.9 &   1.1 & 0.049 & 16.5 & 17.62 &   0.18 & 0.730 & --26.6 & Q      & SBS 1215+497         \\
 12 & 22 & 10.01 & 27 & 19 &  2.3 &   6.4 & 0.046 & 15.9 & 16.90 &   0.52 & 0.442 & --26.3 & Q      & RXS J12221+2719      \\
 12 & 23 &  0.23 & 55 & 40 &  0.3 &   9.4 & 0.057 & 16.5 & 17.48 &   0.48 & 0.905 & --27.5 & Q      & SBS 1220+559         \\
 12 & 28 & 24.97 & 31 & 28 & 37.7 &  24.8 & 0.024 & 15.6 & 16.28 &   0.28 & 2.219 & --30.4 & Q      & B2 1225+31           \\
 12 & 30 & 50.04 &  1 & 15 & 22.6 &   6.4 & 0.041 & 14.4 & 14.81 &   0.16 & 0.117 & --24.8 & Q      & RX J12308+0115       \\
 12 & 33 & 26.05 & 45 & 12 & 23.1 &  11.2 & 0.032 & 16.5 & 16.91 & --0.24 & 1.958 & --29.4 & Q      & HS 1231+4528         \\
 12 & 44 & 10.82 & 17 & 21 &  4.5 &  16.3 & 0.027 & 15.7 & 16.58 &   0.74 & 1.283 & --28.8 & Q      & PG 1241+176          \\
 13 &  6 &  5.72 & 80 &  8 & 20.5 &  10.9 & 0.035 & 16.3 &       &        & 1.183 & --28.3 & Q      & S5 1305+80           \\
 13 & 13 & 21.39 & 78 & 21 & 53.9 &  11.9 & 0.026 & 15.6 &       &        & 2.00  & --30.3 & Q      & HS 1312+7837         \\
 13 & 19 & 56.24 & 27 & 28 &  8.4 &  23.4 & 0.031 & 15.5 & 16.54 &   0.22 & 1.014 & --28.7 & Q      & CSO 873              \\
 13 & 41 &  0.81 & 41 & 23 & 14.2 &  17.0 & 0.022 & 16.4 & 16.36 & --0.02 & 1.204 & --28.3 & Q      & PG 1338+416          \\
 13 & 47 & 19.40 & 59 &  2 & 32.5 &  10.3 & 0.025 & 16.2 & 16.48 &   0.22 & 0.768 & --27.3 & Q      & SBS 1345+592         \\
 13 & 47 & 37.45 & 30 & 12 & 52.4 &  20.5 & 0.035 & 15.7 & 16.12 &   0.68 & 0.118 & --23.6 & Q      & Q J1347+3012         \\
 13 & 51 & 28.30 &  1 &  3 & 38.5 &  17.5 & 0.040 & 16.5 & 17.35 &   0.58 & 1.086 & --27.9 & Q      & Q 1348+0118          \\
 14 &  2 & 44.52 & 15 & 59 & 56.2 &   1.9 & 0.031 & 16.3 & 16.66 &   1.56 & 0.245 & --24.5 & BL     & MC 1400+162          \\
 14 & 27 & 35.60 & 26 & 32 & 14.5 &  26.8 & 0.040 & 15.4 & 15.18 & --0.72 & 0.366 & --26.6 & Q      & PG 1425+267          \\
 14 & 36 & 45.79 & 63 & 36 & 37.6 &  32.1 & 0.038 & 15.7 & 16.61 & --0.22 & 2.066 & --30.4 & Q      & S4 1435+63           \\
 14 & 50 & 26.68 & 58 & 39 & 44.7 &   9.9 & 0.046 & 15.8 & 16.13 &   1.14 & 0.210 & --23.8 & Q      & Mark 830             \\
 14 & 51 & 53.63 & 72 & 14 & 46.8 &   5.8 & 0.027 & 16.3 &       &        & 0.75  & --27.1 & Q      & HS 1451+7227         \\
\hline
\end{tabular}
\end{center}
\end{table*}
\addtocounter{table}{-1}
\begin{table*}
\caption{(end)}
\begin{center}
\begin{tabular}{rrr|rrr|r|c|rrr|l|r|l|l}
\hline \multicolumn{6}{c}{J2000 USNO position} \\
\hline
h & m  & s & $^{\circ}$ & ' & " & " & c/r & O$_{US}$ & O$_{\rm APS}$ & O--E & z & M$_{\rm abs}$ & &  Name  \\
\hline
 15 & 27 & 28.64 & 65 & 48 & 10.3 &  14.0 & 0.046 & 16.2 & 16.77 &   0.56 & 0.345 & --25.5 & Q      & FBS 1526+659         \\
 15 & 50 & 43.65 & 11 & 20 & 47.4 &  23.7 & 0.040 & 16.3 & 16.33 &   0.14 & 0.436 & --25.9 & Q      & MC 1548+114          \\
 15 & 51 & 58.16 & 58 &  6 & 44.7 &   4.5 & 0.036 & 16.2 & 16.74 &   1.14 & 1.320 & --28.7 & Q      & SBS 1550+582         \\
 16 & 19 & 40.54 & 25 & 43 & 23.3 &  13.1 & 0.035 & 16.4 & 16.62 &   1.36 & 0.268 & --24.7 & Q      & RX J16196+2543       \\
 16 & 23 & 19.93 & 41 & 17 &  2.7 &  26.3 & 0.030 & 16.5 & 16.88 &   0.32 & 1.618 & --28.9 & Q      & KUV 16217+4124       \\
 16 & 26 & 37.28 & 58 &  9 & 17.2 &   6.0 & 0.044 & 16.2 &       &        & 0.751 & --27.2 & Q      & SBS 1625+582         \\
 16 & 28 & 25.70 &  8 & 33 &  0.1 &  10.1 & 0.031 & 16.4 & 17.22 &   0.32 & 0.44  & --25.8 & Q      & HS 1626+0839         \\
 16 & 31 & 43.83 & 52 & 53 & 43.8 &   5.7 & 0.048 & 16.5 & 16.81 &   0.44 & 0.352 & --25.2 & Q      & SBS 1630+530         \\
 16 & 32 &  1.11 & 37 & 37 & 50.0 &  14.5 & 0.034 & 16.0 & 16.60 &   0.54 & 1.478 & --29.4 & Q      & PG 1630+377          \\
 16 & 32 & 34.68 & 73 & 59 & 43.0 &   3.2 & 0.030 & 16.1 &       &        & 0.208 & --24.4 & Q      & RXS J16325+7359      \\
 16 & 34 & 29.01 & 70 & 31 & 32.3 &   1.2 & 0.049 & 14.9 &       &        & 1.337 & --30.1 & Q      & PG 1634+706          \\
 16 & 37 &  7.53 & 41 & 40 & 26.9 &  17.0 & 0.028 & 16.2 & 16.28 &   1.60 & 0.765 & --27.3 & Q      & KUV 16355+4146       \\
 17 &  1 &  0.62 & 64 & 12 &  9.3 &  19.0 & 0.035 & 15.9 & 16.03 &   0.18 & 2.736 & --31.2 & Q      & HS 1700+6416         \\
 17 &  4 & 41.39 & 60 & 44 & 30.5 &   8.4 & 0.042 & 15.6 & 15.86 &   0.76 & 0.371 & --26.3 & Q      & 3C 351.0             \\
 17 &  6 & 48.06 & 32 & 14 & 22.9 &  17.0 & 0.029 & 16.2 & 16.94 &   0.66 & 1.070 & --28.2 & Q      & RGB J1706+322        \\
 17 & 19 & 34.18 & 25 & 10 & 58.6 &  16.9 & 0.039 & 16.4 & 16.87 &   1.34 & 0.579 & --26.4 & Q      & RX J17195+2510       \\
 22 & 53 &  7.37 & 19 & 42 & 34.7 &  21.5 & 0.039 & 16.4 & 16.75 &   0.92 & 0.284 & --24.9 & Q      & HS 2250+1926         \\
 23 &  7 & 45.62 & 19 &  1 & 20.8 &  11.6 & 0.033 & 16.4 & 17.13 &   0.98 & 0.313 & --25.0 & Q      & PKS 2305+18          \\
 23 & 11 & 59.50 &  9 & 26 &  1.4 &   7.4 & 0.048 & 16.3 & 17.51 &   0.16 & 0.479 & --26.1 & Q      & RX J23119+0925       \\
 23 & 50 & 10.07 &  8 & 12 & 55.3 &  16.6 & 0.025 & 16.4 & 18.19 &        & 1.70  & --29.1 & Q      & HS 2347+0756         \\
\hline
\end{tabular}
\end{center}
\end{table*}
\normalsize

 Objects brighter than O$_{\rm APS}$=13.5 (zone I) are most probably bright stars.
Indeed among the 576 objects in this zone, 84 could not be classified on the 
Hamburg slitless spectra because of saturation or overlap, but all others are stars 
(25 BA, 416 FG and 51 K).
Objects located above the diagonal line (zone II) are very red and are likely to 
be late-type stars. Fig. \ref{colmag} (c) shows that indeed K and M stars are
mostly located in this zone. Among the 709 objects in this zone, 13 are affected by 
overlapping or saturation, 685 are stars (including 437 K stars); 11 were too weak
to be classified.

\begin{figure}[ht]

\resizebox{4.cm}{!}{\includegraphics{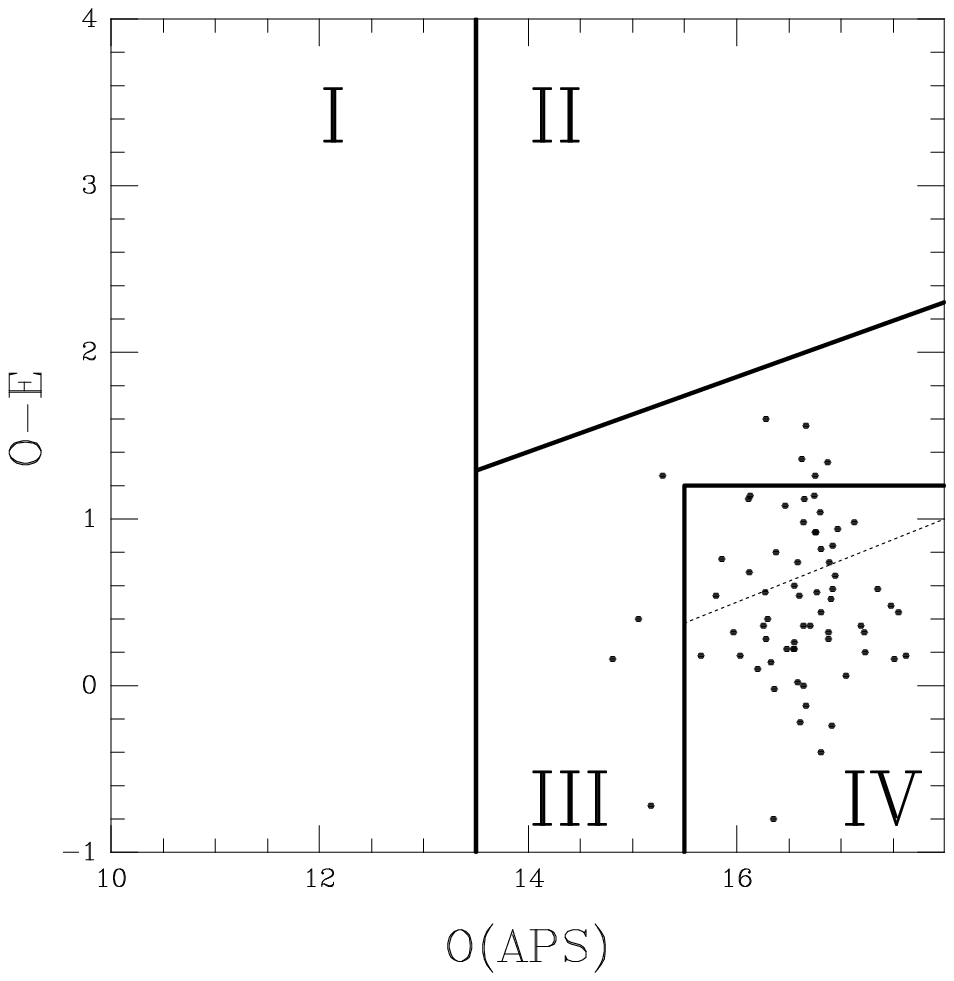}}

\caption{\label{colmag_qso} APS O--E colour indices $\it vs$ O magnitudes for 68
confirmed QSOs.}

\end{figure}

 We cross-correlated the FSC with the current version of the V\'eron-Cetty 
\& V\'eron (\cite{veron01}) AGN catalogue. We found 76 coincidences within 
35$\arcsec$ with a previously known AGN brighter than O$_{\rm USNO}$=16.5 
(they are listed in Table \ref{xrayq}); 68 have magnitude and colour in the 
APS database. Fig. \ref{colmag_qso} is a plot of the APS \mbox{O--E} colour 
index $\it vs$ the O magnitude for these 68 AGN; 87\% (59/68) are located 
in zone IV defined by O$_{\rm APS}$$>$15.5 and O--E$<$1.2. With this limit
on O--E, we exclude red QSOs; this is unavoidable because of the very
large number of stars with O--E$>$1.2. These objects can be recovered
by examination of the slitless spectra. Figs. \ref{colmag} 
(d) and \ref{colmag_qso} show that the distributions of the representative 
points of the EBL-WK and BLUE-WK objects on the one hand and of that of 
QSOs on the other hand are quite similar.   

\begin{figure}[ht]

\resizebox{8.8cm}{!}{\includegraphics{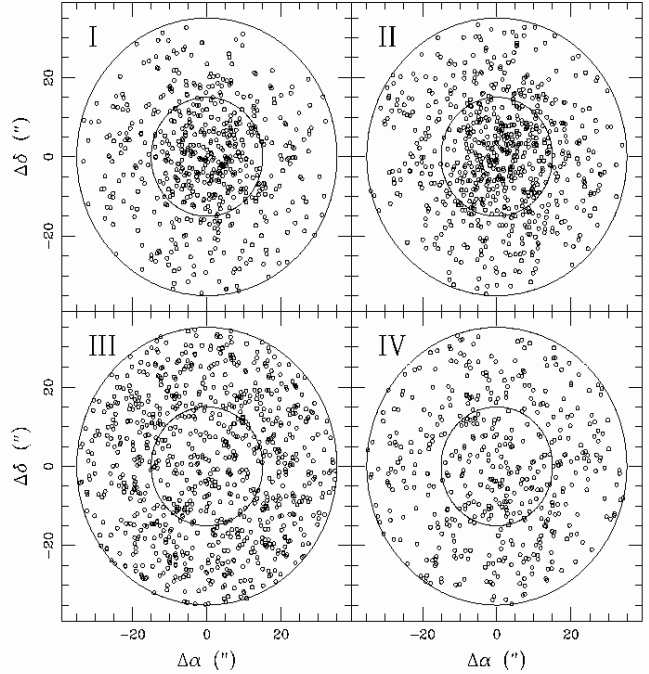}}

\caption{\label{poscol} Position differences between the ROSAT-FSC X-ray sources 
and APS starlike objects located in each of the four zones previously defined. 
The circles have a radius of 15$\arcsec$ and 35$\arcsec$, respectively.}

\end{figure}

 Figure \ref{poscol} shows the differences between the X-ray and optical positions 
for the starlike objects in the 3\,212 sample in each of the four zones defined above. 
In the first two zones there is an obvious concentration towards the centre showing 
that most of the coincidences are genuine associations. In contrast the distribution 
of the points in zone III is quite uniform suggesting that most of these coincidences 
are due to chance.

 In zone III, the observed fraction of coincidences within 15$\arcsec$ is $x$=0.20, 
suggesting that there are indeed very few true associations ($y$=0.05). On the other 
hand, in zone I and II, we found $x$=0.47 and 0.45 respectively showing that 
basically all objects (90 and 84\% respectively) are real associations. 

 For the 212 starlike objects classified as BLUE-WK on the Hamburg slitless spectra, 
we found $x$=0.24, so the fraction of true coincidences is about 18\%, while for 
the 84 EBL-WK objects, $x$=0.51 and, within the statistical uncertainties, all are 
genuine associations.

\begin{table*}
\caption{\label{catalog} List of 103 QSO candidates identified with a ROSAT-FSC source
and located in zone IVa.
Col. 1: right ascension, col. 2: declination, col. 3: separation between 
the ROSAT and USNO positions in arcsec, col. 4: X-ray count 
rate (count s$^{-1}$), col. 5: magnitude O$_{\rm USNO}$, col. 6: magnitude O$_{\rm APS}$, 
col. 7: APS O--E colour, col. 8: redshift, col. 9: absolute O$_{\rm USNO}$ magnitude, 
col. 10: classification: DA, DN, WD: white dwarfs; *: stars; CV: cataclysmic variable; 
BL: BL\,Lac object; S1: Seyfert 1 galaxy; S1n: Narrow Line Seyfert 1; Q: QSO; 
col. 11: reference: (1) OAGH and (2) OHP observations, col. 12: name.}
\begin{center}
\begin{tabular}{rrr|rrr|r|c|crr|c|l|l|c|l}
\hline \multicolumn{6}{c}{J2000 USNO position} \\
\hline
h & m  & s & $^{\circ}$ & ' & " & " & c/r & O$_{\rm US}$ & O$_{\rm APS}$ & O--E & z & M$_{\rm abs}$ & & Ref. & Name  \\
\hline

  0 & 11 & 19.46 & 28 & 17 & 50.8 &  18.7 & 0.023 & 15.5 & 15.88 &   0.12 &       &        & *     & 1 &                      \\
  0 & 13 &  4.80 & 10 & 11 & 28.6 &   5.8 & 0.023 & 16.0 & 16.25 &   0.32 & 0.239 & --24.9 & Q     & 1 &                      \\
  0 & 40 & 46.17 & 18 & 54 & 24.3 &  17.0 & 0.026 & 16.0 & 16.71 & --0.12 & 0.153 & --23.3 & Q     & 1 &                      \\
  1 &  7 & 37.51 & 22 & 22 & 32.3 &  33.7 & 0.042 & 16.2 & 16.89 &   0.68 &       &        & *     & 2 &                      \\
  1 & 20 & 46.96 & 23 & 31 & 40.1 &  28.7 & 0.047 & 15.9 & 16.27 &   0.54 &       &        & *     & 2 &                      \\
  1 & 37 & 37.24 & 30 &  2 & 49.1 &  13.1 & 0.035 & 13.5 & 16.90 &   0.36 &       &        & DN    &   & TX Tri               \\
  2 & 55 & 30.70 &  7 & 25 & 57.1 &  19.4 & 0.030 & 16.3 & 16.97 &   0.66 & 0.707 & --27.0 & Q     & 1 &                      \\
  3 & 53 & 15.60 &  9 & 56 & 35.1 &  24.0 & 0.030 & 16.0 & 16.49 & --0.82 &       &        & *     & 2 &                      \\
  8 & 23 & 46.65 & 24 & 53 & 51.2 &  18.5 & 0.039 & 15.6 & 15.83 & --0.40 &       &        & WD    & 1 & TON  316             \\
  8 & 39 & 34.28 & 23 & 34 & 10.4 &  12.8 & 0.041 & 15.8 & 16.49 & --0.52 &       &        & DA    &   & PG 0836+237          \\
  8 & 45 & 51.14 & 60 &  9 & 13.7 &  13.3 & 0.027 & 16.0 & 16.30 & --0.46 &       &        & DA    &   & PG 0841+603          \\
  8 & 50 & 20.09 & 54 & 33 & 50.5 &  17.3 & 0.038 & 16.4 & 16.78 &   0.62 &       &        & *     & 1 &                      \\
  8 & 56 & 32.44 & 50 & 41 & 14.1 &   5.1 & 0.027 & 15.8 & 15.57 & --0.58 & 0.235 & --25.0 & Q     & 1 &                      \\
  9 &  3 & 32.62 & 16 & 26 &  1.4 &  21.7 & 0.037 & 16.2 & 16.59 &   0.24 &       & & *     & 1 & LB 9181              \\
  9 & 10 & 35.38 & 31 & 27 & 45.9 &  27.0 & 0.048 & 16.3 & 16.21 &   0.42 &       & & *     & 1 &                      \\
  9 & 10 & 37.97 & 53 & 12 &  2.8 &  19.3 & 0.030 & 16.4 & 16.82 &   0.52 &       & & *     & 1 &                      \\
  9 & 12 & 15.50 &  1 & 19 & 59.3 &   3.8 & 0.041 & 16.5 & 16.76 & --0.50 &       & & *     & 1 &                      \\
  9 & 20 & 47.61 & 51 & 25 & 38.8 &  33.1 & 0.021 & 15.7 & 15.65 &   0.12 &       & & *     & 1 &                      \\
  9 & 30 &  6.76 & 52 & 28 &  4.0 &  12.0 & 0.039 & 15.8 & 16.37 & --0.02 &       & & *     &   & PG 0926+527          \\
  9 & 30 & 21.87 & 23 & 53 & 30.1 &  19.2 & 0.027 & 16.0 & 16.70 &   0.14 & 0.243 & --24.3 & Q     & 1 &                      \\
  9 & 30 & 34.46 & 20 & 44 & 16.9 &   5.2 & 0.045 & 16.1 & 17.09 &   0.64 & 1.169 & --28.4 & Q     & 1 &                      \\
  9 & 45 & 13.97 & 15 & 10 & 11.7 &  17.9 & 0.057 & 15.8 & 17.66 &   0.86 &       & & *     & 1 &                      \\
  9 & 46 & 34.50 & 13 & 50 & 58.3 &  34.4 & 0.027 & 15.8 & 16.75 &   0.10 &       & & CV    & 1 &                      \\
  9 & 49 & 39.77 & 17 & 52 & 49.5 &  12.0 & 0.021 & 16.3 & 18.65 &   0.60 &       & &       &   &                      \\
  9 & 52 & 45.70 &  2 &  9 & 38.7 &  17.1 & 0.025 & 15.2 & 16.32 & --0.76 &       & & DA    &   & PG 0950+024          \\
  9 & 56 & 49.88 & 29 & 50 & 14.6 &  10.7 & 0.046 & 15.4 & 16.09 & --0.10 & 0.845 & --28.5 & Q     & 1 & TON  465             \\
  9 & 57 & 11.78 & 63 & 10 & 10.2 &   3.8 & 0.024 & 16.4 & 16.25 & --0.06 & 0.918 & --27.6 & Q     & 1 &                      \\
 10 &  6 & 14.64 & 44 & 19 &  7.1 &  15.6 & 0.030 & 16.5 & 16.93 & --0.30 &       & & DA    &   &                      \\
 10 & 34 & 53.07 & 44 & 57 & 23.1 &  20.5 & 0.023 & 16.1 & 16.57 &   0.38 & 1.422 & --29.0 & Q     & 1 &                      \\
 10 & 35 & 27.50 & 49 & 58 & 27.7 &  22.8 & 0.021 & 16.5 & 17.44 &   0.36 & 1.427 & --28.5 & Q     & 1 &                      \\
 10 & 44 & 19.33 & 19 & 57 & 47.9 &  24.2 & 0.039 & 15.6 & 15.90 &   0.44 &       & & *     & 1 &                      \\
 10 & 46 & 42.30 & 39 & 20 & 18.2 &   8.3 & 0.043 & 16.4 & 17.16 &   0.22 & 0.390 & --25.3 & Q     & 1 &                      \\
 10 & 47 & 30.55 & 10 & 17 & 28.9 &  17.7 & 0.228 & 15.4 & 15.81 &   0.08 & 0.145 & --24.3 & Q     & 1 &                      \\
 11 &  8 & 42.46 & 16 & 50 & 40.2 &  31.2 & 0.030 & 15.6 & 16.48 &   0.04 &       & & *     & 1 &                      \\
 11 & 14 &  1.95 & 52 & 27 & 11.6 &  29.8 & 0.023 & 15.5 & 15.62 &   0.40 &       & & *     & 1 &                      \\
 11 & 15 &  7.70 &  2 & 37 & 57.7 &   7.6 & 0.024 & 16.5 & 17.08 & --0.04 & 0.564 & --26.1 & Q     & 1 &                      \\
 11 & 26 & 16.19 & 32 & 59 & 53.7 &  29.3 & 0.029 & 16.2 & 15.66 &   0.24 &       & & *     & 1 &                      \\
 11 & 33 & 31.23 & 58 & 57 & 47.8 &  19.9 & 0.025 & 15.9 & 15.98 &   0.44 &       & & *     & 1 &                      \\
 11 & 34 & 25.08 & 23 & 16 &  8.8 &  24.0 & 0.039 & 16.3 & 16.66 &   0.64 &       & & *     & 1 &                      \\
 11 & 38 & 36.32 & 47 & 55 & 10.0 &  13.3 & 0.046 & 16.1 & 17.53 & --0.48 &       & & *     & 1 &                      \\
 11 & 41 & 52.82 & 25 & 35 & 33.5 &  10.8 & 0.042 & 16.3 & 16.82 & --0.96 &       & & *     & 1 &                      \\
 11 & 47 & 47.34 & 26 &  0 & 49.0 &  10.9 & 0.042 & 15.5 & 15.84 &   0.00 &       & & *     & 1 &                      \\
 12 & 19 & 57.97 & 27 &  8 & 57.2 &   9.5 & 0.023 & 16.1 & 17.17 &   0.76 &       & & *     & 1 &                      \\
 12 & 31 & 25.72 & 25 & 55 & 59.8 &   5.7 & 0.032 & 15.8 & 16.16 &   0.54 &       & & *     & 1 &                      \\
 12 & 32 & 54.28 & 36 & 44 &  7.4 &  32.5 & 0.022 & 16.0 & 16.61 & --0.08 &       & & WD    & 1 & CBS  353             \\
 12 & 52 & 30.85 & 14 & 26 &  9.3 &  24.3 & 0.029 & 16.3 & 17.30 &   0.10 &       & &       &   &                      \\
 12 & 57 & 37.02 & 16 & 30 & 48.6 &  13.5 & 0.035 & 16.1 & 16.95 &   0.66 & 1.017 & --28.2 & Q     & 1 &                      \\
 12 & 59 & 38.22 & 60 & 38 & 59.4 &  15.9 & 0.039 & 16.3 & 16.47 & --0.48 &       & & *     & 1 & SBS 1257+609         \\
 12 & 59 & 44.45 & 68 &  4 &  0.8 &  30.9 & 0.022 & 16.3 & 16.64 & --0.72 &       & & *     & 1 & FBS 1257+683         \\
 13 &  0 &  6.40 & 44 & 42 & 50.9 &  13.9 & 0.036 & 16.2 & 16.06 &   0.46 &       & & *     & 1 &                      \\
 13 & 10 & 11.30 &  7 & 58 & 16.5 &  21.2 & 0.030 & 15.8 & 16.52 & --0.08 & 0.578 & --27.0 & Q     & 1 &                      \\
 13 & 13 & 15.90 &  9 & 18 & 20.4 &  27.7 & 0.021 & 16.4 & 17.34 &   0.10 & 1.790 & --29.2 & Q     & 1 &                      \\
 13 & 14 & 30.78 & 13 &  7 & 45.8 &  23.8 & 0.044 & 16.3 & 17.00 &   0.24 & 0.741 & --27.1 & Q     & 1 &                      \\
 13 & 20 &  1.10 &  7 & 18 & 17.1 &  21.1 & 0.038 & 16.5 & 16.97 &   0.74 & 0.866 & --26.9 & Q     & 1 &                      \\
 13 & 20 & 22.53 & 30 & 56 & 22.4 &  19.8 & 0.029 & 16.4 & 17.22 & --0.04 & 1.587 & --28.1 & Q     & 1 & US  583              \\
 13 & 24 & 47.70 &  3 & 24 & 32.9 &  15.5 & 0.040 & 16.5 & 16.84 &   0.54 & 0.303 & --24.9 & Q     & 1 &                      \\
\hline
\end{tabular}
\end{center}
\end{table*}
\addtocounter{table}{-1}
\begin{table*}
\caption{(end)}
\begin{center}
\begin{tabular}{rrr|rrr|r|c|crr|c|l|l|c|l}
\hline \multicolumn{6}{c}{J2000 USNO position} \\
\hline
h & m  & s & $^{\circ}$ & ' & " & " & c/r & O$_{US}$ & O$_{\rm APS}$ & O--E & z & M$_{\rm abs}$ & & Ref. & Name  \\
\hline
 13 & 32 & 51.09 & 15 & 29 & 30.9 &  23.9 & 0.042 & 15.5 & 15.56 &   0.14 &       & & *     & 1 &                      \\
 13 & 33 & 45.80 &  5 & 38 & 42.6 &  22.2 & 0.028 & 15.5 & 15.74 &   0.10 &       & & *     & 1 &                      \\
 13 & 40 & 59.95 & 60 & 26 & 11.8 &  17.6 & 0.034 & 16.2 & 16.48 & --0.68 &       & & DA    &   & SBS 1339+606         \\
 13 & 42 & 46.90 & 18 & 44 & 43.8 &  22.4 & 0.039 & 16.4 & 16.69 &   0.12 & 0.382 & --25.5 & Q     & 1 &                      \\
 13 & 50 & 13.97 & 14 & 35 & 47.4 &  30.4 & 0.023 & 16.4 & 16.93 &   0.48 &       & & *     & 1 &                      \\
 13 & 51 & 25.25 & 19 &  5 & 33.4 &  10.8 & 0.027 & 16.4 & 16.89 &   0.64 &       & & *     & 1 &                      \\
 13 & 54 & 48.82 & 49 & 13 & 37.1 &  19.8 & 0.028 & 15.5 & 15.68 &   0.32 &       & & *     & 1 &                      \\
 13 & 55 & 43.99 & 20 & 12 & 31.8 &  19.9 & 0.030 & 15.6 & 16.11 &   0.30 &       & & *     & 1 &                      \\
 13 & 58 & 41.50 &  2 & 49 & 12.5 &  21.1 & 0.046 & 16.2 & 16.85 & --0.08 &       & & *     & 1 &                      \\
 14 &  6 & 58.74 & 14 & 42 & 38.6 &  13.0 & 0.034 & 15.4 & 15.84 &   0.34 & 0.264 & --25.7 & Q     & 1 &                      \\
 14 &  9 & 39.24 & 28 & 16 & 49.9 &  25.2 & 0.035 & 16.1 & 16.93 &   0.34 & 0.165 & --23.9 & Q     & 1 &                      \\
 14 & 10 & 57.74 & 64 & 33 & 10.6 &  18.8 & 0.041 & 16.4 & 16.83 &   0.06 & 0.462 & --25.9 & Q     & 1 &                      \\
 14 & 17 & 30.16 & 13 &  0 &  1.6 &  11.6 & 0.021 & 15.7 & 16.00 &   0.38 &       & & *     & 1 &                      \\
 14 & 19 & 25.77 & 38 &  2 & 49.0 &  12.7 & 0.039 & 16.2 & 17.22 &   0.08 & 0.517 & --26.3 & Q     & 1 &                      \\
 14 & 31 & 10.98 & 14 & 23 &  8.3 &   3.7 & 0.022 & 16.1 & 15.91 & --0.04 & 1.425 & --29.0 & Q     & 1 &                      \\
 14 & 47 & 50.12 & 38 &  5 & 30.3 &  12.2 & 0.023 & 15.8 & 16.28 &   0.04 &       & & *     & 1 &                      \\
 15 &  0 & 31.80 & 48 & 36 & 47.0 &  16.0 & 0.021 & 16.4 & 16.72 &   0.48 & 1.031 & --27.9 & Q     & 1 &                      \\
 15 &  5 & 27.62 & 29 & 47 & 18.7 &   8.7 & 0.036 & 15.0 & 15.65 & --0.34 & 0.527 & --27.6 & Q     & 1 & CSO 1080             \\
 15 &  8 & 32.28 & 67 & 42 & 43.4 &   9.0 & 0.032 & 16.0 & 17.12 &   0.14 & 0.336 & --25.8 & Q     & 1 &                      \\
 15 & 12 & 44.60 &  9 & 31 &  0.8 &  23.0 & 0.037 & 15.9 & 16.61 &   0.56 &       & & *     & 1 &                      \\
 15 & 16 & 32.30 & 12 & 13 & 50.5 &  10.6 & 0.028 & 16.2 & 17.26 &   0.78 &       & & BL    & 1 &                      \\
 15 & 44 &  3.76 & 26 & 48 & 38.6 &  21.7 & 0.023 & 16.4 & 17.15 &   0.74 &       & & *     & 1 &                      \\
 15 & 45 & 53.50 &  9 & 36 & 20.6 &  19.6 & 0.032 & 16.1 & 16.40 & --0.22 &       & &       &   &                      \\
 15 & 48 & 33.03 & 44 & 22 & 26.1 &  12.7 & 0.021 & 16.3 & 16.97 &   0.48 & 0.322 & --25.2 & Q     & 1 &                      \\
 15 & 51 &  9.60 & 45 & 42 & 52.1 &  22.5 & 0.036 & 15.6 & 15.95 &   0.32 &       & &       &   &                      \\
 15 & 51 & 52.45 & 19 & 11 &  4.1 &  34.9 & 0.034 & 15.8 & 16.31 &   0.50 &       & &       &   &                      \\
 15 & 53 &  4.93 & 35 & 48 & 28.6 &  25.2 & 0.033 & 16.1 & 17.20 &   0.72 & 0.722 & --26.8 & Q     & 1 &                      \\
 15 & 56 &  9.90 &  3 &  9 & 22.5 &   9.6 & 0.037 & 16.3 & 16.80 &   0.48 & 0.131 & --23.2 & S1n   & 1 &                      \\
 16 &  5 & 19.72 & 14 & 48 & 52.5 &  15.5 & 0.040 & 16.2 & 16.51 &   0.00 & 0.371 & --25.7 & Q     & 1 &                      \\
 16 &  9 & 47.80 &  7 & 12 & 33.2 &  27.8 & 0.021 & 16.4 & 16.95 &   0.64 &       & & *     & 1 &                      \\
 16 & 11 & 36.58 & 15 & 20 & 54.6 &  18.8 & 0.030 & 16.3 & 16.69 &   0.40 & 1.309 & --28.6 & Q     & 1 &                      \\
 16 & 17 & 42.10 &  6 &  2 & 23.8 &  20.4 & 0.078 & 15.5 & 15.69 &   0.42 &       & &*     & 1 &                      \\
 16 & 28 & 35.38 & 45 & 20 & 43.3 &  34.1 & 0.022 & 16.4 & 17.00 &   0.68 &       & & *     & 1 &                      \\
 16 & 30 &  9.65 & 45 & 16 & 37.9 &  21.7 & 0.022 & 16.5 & 16.96 &   0.32 &       & & *     & 1 &                      \\
 16 & 31 & 10.46 & 38 & 44 & 49.6 &   7.8 & 0.036 & 15.8 & 15.97 &   0.48 &       & & *     & 1 &                      \\
 16 & 45 & 20.15 & 61 & 35 &  9.7 &   3.8 & 0.023 & 16.4 & 16.89 &   0.64 & 0.410 & --25.5 & Q     & 1 &                      \\
 16 & 46 & 15.52 & 25 & 41 & 43.3 &  15.7 & 0.047 & 16.4 & 17.26 &   0.56 & 0.188 & --23.9 & Q     & 1 &                      \\
 17 & 18 & 28.97 & 57 & 34 & 22.3 &   6.4 & 0.039 & 15.9 & 16.07 &   0.48 & 0.100 & --22.9 & S1n   & 1 &                      \\
 17 & 20 & 13.19 & 49 & 55 & 26.4 &  28.8 & 0.027 & 16.0 & 15.89 &   0.36 &       & & *     & 1 &                      \\
 17 & 21 & 45.75 & 57 & 16 & 56.7 &  31.5 & 0.021 & 15.8 & 15.87 &   0.24 &       & & *     & 1 &                      \\
 17 & 28 & 13.44 & 32 & 22 &  5.8 &  10.8 & 0.022 & 16.4 & 17.08 &   0.34 & 0.563 & --26.3 & Q     & 1 &                      \\
 17 & 29 & 35.53 & 52 & 30 & 47.5 &   3.0 & 0.029 & 16.2 & 15.96 &   0.02 & 0.278 & --25.0 & Q     & 1 &                      \\
 17 & 49 & 12.30 & 55 & 12 &  8.9 &  28.7 & 0.023 & 16.3 & 16.58 &   0.44 &       & & WD    & 1 &                      \\
 22 & 37 & 31.84 & 10 & 19 &  4.0 &   3.8 & 0.024 & 16.5 & 16.77 & --0.46 & 0.103 & --22.5 & S1    & 1 &                      \\
 22 & 37 & 33.28 & 10 & 18 & 42.7 &  26.3 & 0.024 & 15.9 & 16.10 &   0.50 &       & & *     & 1 &                      \\
 23 &  7 & 13.30 &  4 & 32 &  2.5 &  29.8 & 0.079 & 15.9 & 17.25 & --0.92 &       & & *     & 1 &                      \\
 23 & 15 & 52.73 & 11 & 33 &  2.1 &  21.4 & 0.036 & 16.5 & 17.20 &   0.24 & 0.567 & --26.2 & Q     & 1 & KUV 23134+1117       \\
\hline
\end{tabular}
\end{center}
\end{table*}
\normalsize

\begin{figure}[ht]

\resizebox{8.8cm}{!}{\includegraphics{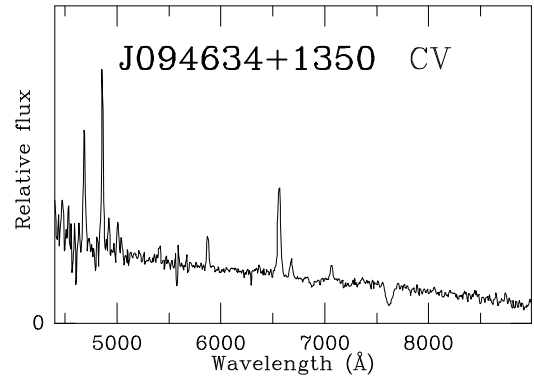}}

\caption{\label{sp_raul_cv} Spectrum of a cataclysmic variable observed at OAGH.}

\end{figure}

\begin{figure}[ht]

\resizebox{8.8cm}{!}{\includegraphics{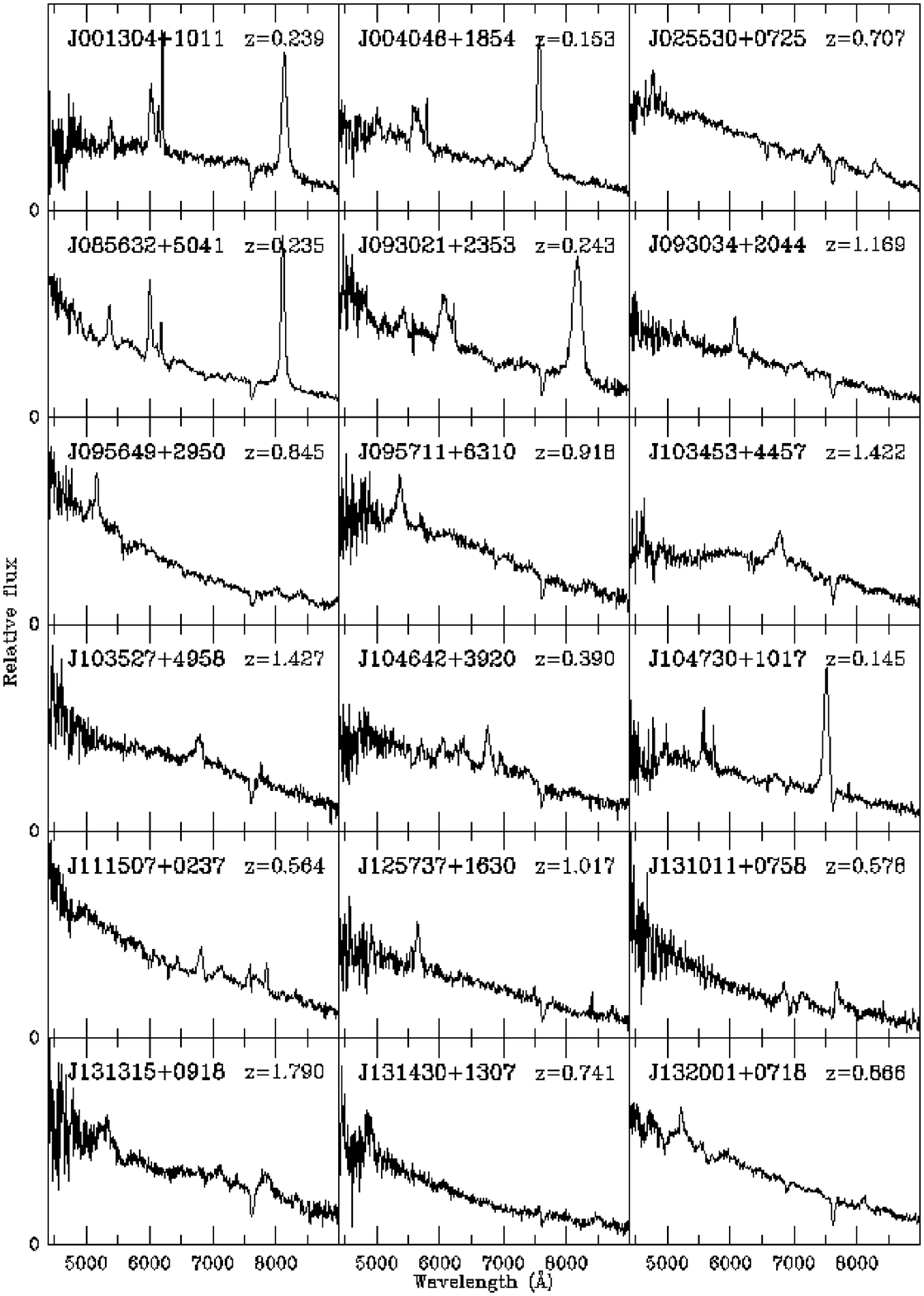}}

\caption{\label{sp_raul_1} Spectra of 18 QSOs from Table 4 observed at OAGH.}

\end{figure}

\begin{figure}[ht]

\resizebox{8.8cm}{!}{\includegraphics{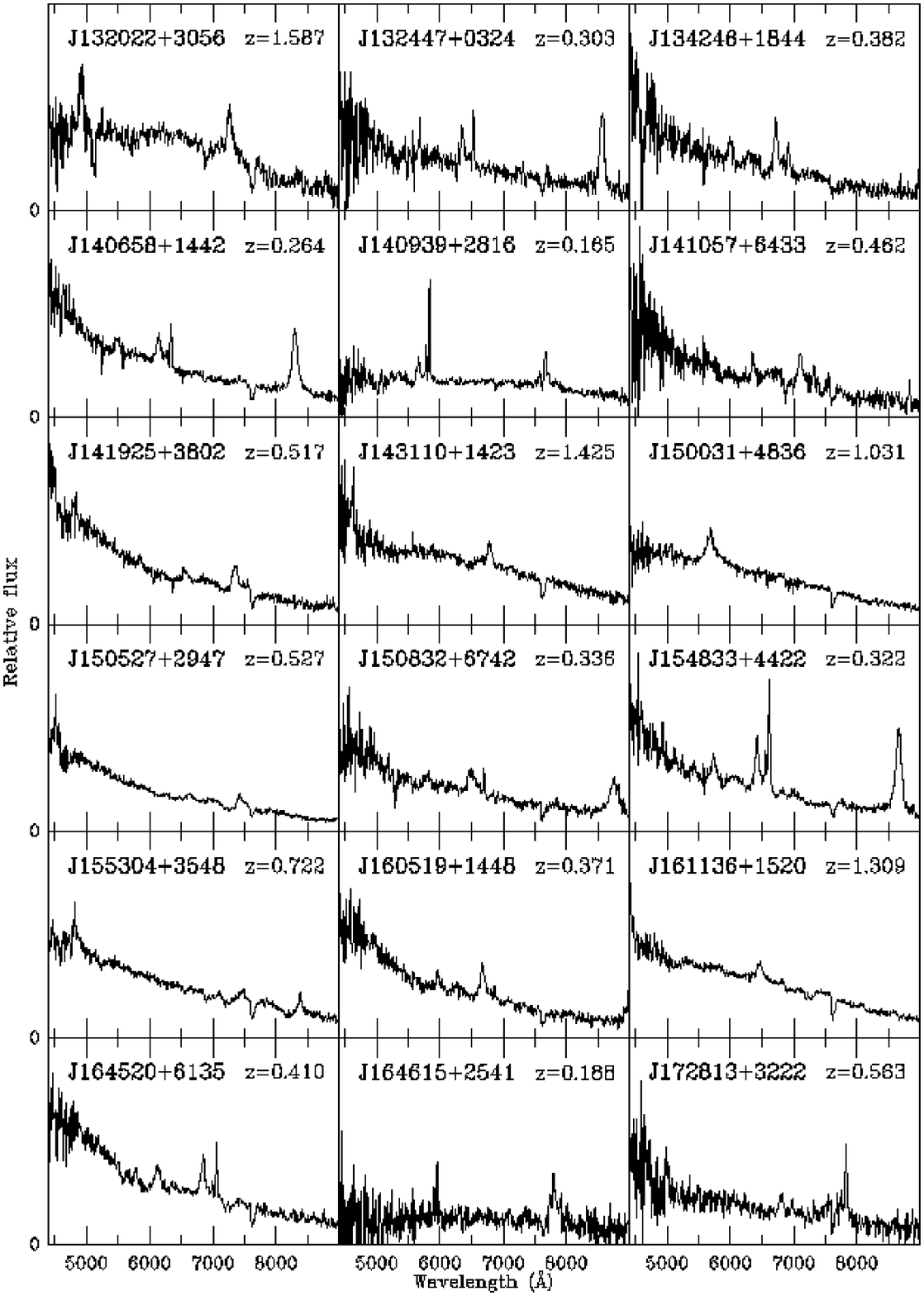}}

\caption{\label{sp_raul_2} Spectra of 18 QSOs from Table 4 observed at OAGH.}

\end{figure}

\begin{figure}[ht]

\resizebox{8.8cm}{!}{\includegraphics{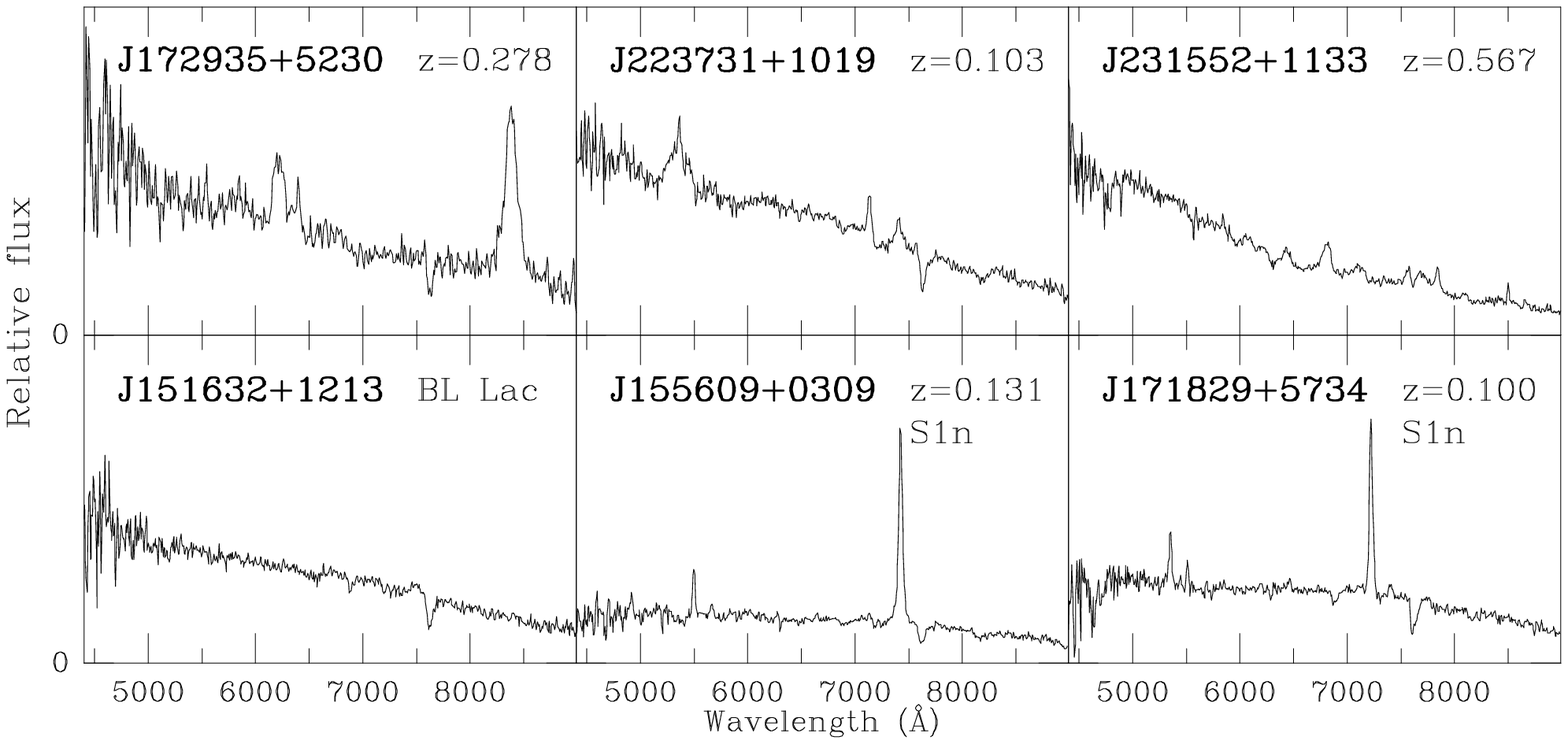}}

\caption{\label{sp_raul_3} Spectra of six objects from Table 4 (one Seyfert 1, one QSO, 
two BL~Lac objects and two NLS1s) observed at OAGH.}

\end{figure}

 Out of the 2\,652 starlike objects, zone IV contains 476 coincidences within 
35$\arcsec$ and 136 (x=0.28) within 15$\arcsec$ of an X-ray source, implying 
that y=0.32; therefore 153 of these coincidences are likely 
to be genuine associations, 76 being located within 15$\arcsec$. They could be 
blue stars (WD, CV, sd), or more probably QSOs. We have divided zone IV in two 
subregions (see the dotted line on Fig. \ref{colmag}). In the upper region (IVb), 
only 23\% of the coincidences are within 15$\arcsec$ and therefore a small 
fraction of the 332 coincidences are genuine. In the lower region (IVa) 
containing the bluest objects, 41\% of the coincidences are located within 
15$\arcsec$ from the X-ray source, suggesting that $\sim$70\% of the 144 objects 
located in this zone are real identifications. Among them are 41 known QSOs. The 
103 other candidates, of which seven are known stars, are listed in Table 
\ref{catalog}. We have spectroscopically observed 91 of them. The observations 
were carried out during three observing runs (May and August 2002 and May 2003)
with the 2.1-m telescope of the Guillermo Haro Astrophysical Observatory (OAGH) 
located in Cananea (Sonora, Mexico), operated by INAOE, and one run (September 
2003) with the OHP 1.93-m telescope and the same instrument setting as for the 
observations of the galaxies (see section 3). The 2.1-m telescope was equipped 
with the LFOSC focal reducer (Zickgraf et al. \cite{zickgraf97}). The slit mask 
and the lower dispersion grism were used, giving a wavelength coverage from 4200 
\AA\ to 9000 \AA\ and a dispersion of 8.2 \AA\ pix$^{-1}$. We found 49 stars 
(including a CV the spectrum of which is shown in Fig. \ref{sp_raul_cv}), 38 QSOs, 
1 Seyfert 1, two NLS1s, one BL~Lac object (their spectra are displayed in Figs. 
\ref{sp_raul_1}, \ref{sp_raul_2} and \ref{sp_raul_3}). Five candidates have not 
yet been observed. 

 Altogether, there are, in region IVa, 83 confirmed AGN in a total of 139 
spectroscopically observed starlike objects (60\%).


 But only 60\% (41/68) of all previously known AGN with APS colour fall in zone IVa 
showing the limits 
of the efficiency of the procedure we have adopted. The others are located in zone IVb, 
or not far from it, in a region where chance coincidences between stars and X-ray sources 
become non negligible. Here the Hamburg slitless spectra may turn out to 
be most useful in separating AGN from stars.

\section{Discussion}

\subsection{Identification procedure}

 We have seen that the galaxies found within 35$\arcsec$ of a ROSAT source have a 
high probability of being genuine identifications ($\sim$80\%) and therefore we 
included all of them in our spectroscopic program.
  
 In the case of starlike objects, the situation is more complex due to the facts that 
a large fraction of the X-ray sources are in fact X-ray stars and that, the surface 
density of stars being relatively large, there is a non negligible number of chance 
coincidences.
 The use of the O--E {\it vs.} O diagram turns out to be quite efficient to eliminate 
bright stars and late type stars. However, although in our region IVa most objects are 
genuine identifications, in zones IVb and III there are many chance coincidences with 
main sequence stars. The use of the Hamburg slitless spectra in these zones is crucial 
to separate QSOs from stars. In addition, Seyfert 1 galaxies are often compact and 
classified as starlike in the APS. Moreover they usually have a relatively red O--E 
colour; as a consequence they easily escape detection by our automatic procedure and
classification on the basis of the slitless spectra is necessary.

 The relatively poor positional accuracy of the ROSAT survey ($\sigma$$\sim$15$\arcsec$) 
makes it difficult to identify objects weaker than $\sim$16.5; but this limiting 
magnitude is well fitted by the magnitude limit below which no recognition of 
absorption features is possible in the Hamburg slitless spectra.

\subsection{Completeness of our QSO sample}

  We have seen that only 85\% of the bright USNO objects have a counterpart in the APS
due mainly to the incompleteness of the data base. Moreover only 60\% of the known 
QSOs fall into our zone IVa to which we have restricted our search. Therefore our 
sample is expected to be no more than 50\% complete. \\

  On the other hand, as our aim is to find all AGN brighter than a given optical 
magnitude, we have to answer the following question: how many such AGN are missed 
if we identify all bright AGN in a flux limited X-ray survey? \\

\begin{table}[ht]
\caption{\label{xopt} Number and percentage of radio quiet and radio loud X-ray 
detected QSOs in the V\'eron-Cetty \& V\'eron catalogue. Cols. 1 and 2: O$_{\rm USNO}$ 
range, col. 3: number of QSOs, col. 4: number of sources brighter than 0.02 count 
s$^{-1}$ in the ROSAT catalogues, col. 5: percentage of ROSAT detected QSOs; col. 
6 to 8: the same for the radio loud QSOs.}
\begin{center}
\begin{tabular}{rr|rrr|rrr}
\hline  \multicolumn{2}{c}{} & \multicolumn{3}{c}{Radio quiet} & \multicolumn{3}{c}{Radio loud} \\ 
\hline
O$_{min}$ & O$_{max}$  & n & n$_{X}$ &  \% & n & n$_{X}$ & \%  \\
\hline
      &$<$15.5 &   80 &   58 & 72 &   9 &  7 & 78 \\
 15.5 &   16.0 &   90 &   48 & 53 &  21 & 17 & 81 \\
 16.0 &   16.5 &  243 &  101 & 42 &  49 & 37 & 76 \\
 16.5 &   17.0 &  433 &  144 & 33 &  99 & 60 & 61 \\
 17.0 &   17.5 &  679 &  101 & 15 & 172 & 77 & 45 \\
 17.5 &   18.0 &  900 &   60 &  7 & 201 & 59 & 29 \\
 18.0 &   18.5 &  977 &   23 &  2 & 169 & 41 & 24 \\
 18.5 &   19.0 & 1040 &   13 &  1 & 107 & 11 & 10 \\
 19.0 &   19.5 &  475 &    4 &  1 &  41 &  5 & 12 \\
 19.5 &   20.0 &   87 &    0 &  0 &  18 &  5 & 28 \\
\hline
\end{tabular}
\end{center}
\end{table}
\normalsize

 The current version of the V\'eron \& V\'eron QSO catalogue contains 5\,893 
QSOs at $\delta>$0$^{\circ}$, $\vert$b$\vert$$>$30$^{\circ}$, with B$<$20.0 and 
z$>$0.10, found in the course of radio or optical surveys (QSOs 
found as the identification of a X-ray source have been ignored). Of these, 871 
are located within 35$\arcsec$ from a ROSAT-BSC or -FSC source (stronger than 0.02 
count s$^{-1}$). We consider them as genuine identifications; 319 are radio loud 
and 552 radio quiet. 

\begin{figure}[ht]

\resizebox{5cm}{!}{\includegraphics{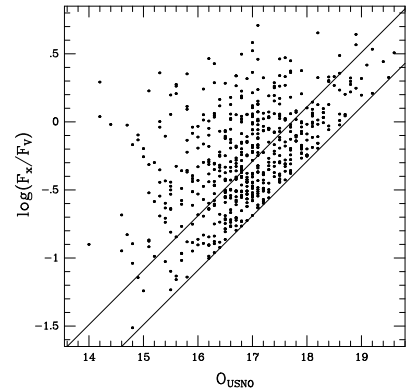}}

\caption{\label{fxfopt} Plot of log(F$_{X}$/F$_{V}$) $\it vs$ O$_{\rm USNO}$ for 
552 radio quiet QSOs. The two diagonal lines correspond to the sensitivity limits 
of the BSC and FSC ROSAT survey (0.05 and 0.02 count s$^{-1}$, respectively).}
\end{figure}

 On average, QSOs with higher optical luminosity have higher X-ray luminosity. 
According to Maccacaro et al. (\cite{maccacaro88}),
log(F$_{X}$/F$_{V}$)=log(F$_{X}$)+0.4$\times$V+5.37 where F$_{X}$ is the X-ray flux in 
the 0.3-3.5 keV band in erg cm$^{-2}$ s$^{-1}$ and F$_{V}$=--0.4$\times$V--8.42 is
the flux in the V band in the same units. For the ROSAT survey, this translates 
into: log(F$_{X}$/F$_{V}$)=log(PSPC counts s$^{-1}\times$10$^{-11}$)+0.4$\times$V+5.37
where F$_{X}$ is the X-ray flux in the 0.1-2.4 keV band, assuming a conversion 
factor of 1 PSPC count s$^{-1}$ for a flux of 10$^{-11}$ erg cm$^{-2}$ s$^{-1}$ in this 
band which corresponds to an unabsorbed source with a photon index $\Gamma$=2.3, a
typical value for QSOs (Voges et al. \cite{voges99}). 
 
 Fig. \ref{fxfopt} shows log(F$_{X}$/F$_{V}$) $\it vs$ O$_{\rm USNO}$ 
for the 552 radio quiet QSOs. Table \ref{xopt} gives the fraction of ROSAT 
detected QSOs ${\it vs}$ their optical magnitude for both the radio quiet and 
the radio loud samples. 

 At O$_{\rm USNO}$=16.0, about half of the optically selected QSOs are not 
detected in X-ray (Table \ref{xopt}) and therefore have log(F$_{X}$/F$_{V}$)$<$--1 . 
Consequently the median value of their X-ray to optical flux ratio is 
\mbox{$\sim$--1.0}. Figure \ref{fxfopt} suggests that $\sigma$$\sim$0.5. 

 Zickgraf et al. (\cite{zickgraf03}) found that the median value of 
log(F$_{X}$/F$_{V}$) for X-ray selected QSOs is 0.39. The fact that these 
objects are significantly stronger X-ray sources than optically 
selected QSOs is due to the much higher frequency of the survey.
Indeed Williams \& Bridle (\cite{williams67}) and Kellermann et al. 
(\cite{kellermann68}) have shown that the relative numbers of sources with 
given spectral indices depend on the frequency at which the sample is 
selected. A sample selected at a low frequency (optical) will include a greater 
proportion of sources with high values of the spectral index than will a similar
sample selected at a high frequency (X-ray).

 In the simple case where the distribution of the spectral indices $\alpha$ in
a complete sample of sources selected at a wavelength $\lambda$$_{o}$ is Gaussian with
a mean spectral index $\alpha$$_{o}$ and a standard deviation $\sigma$$_{\alpha}$,
the distribution of the spectral indices of sources selected at another wavelength
$\lambda$$_{\rm X}$ will also be Gaussian with the same standard deviation 
$\sigma$$_{\alpha}$, but with a mean spectral index $\alpha$$_{\rm X}$ given by:

\mbox{$\alpha$$_{\rm X}$=$\alpha$$_{o}$+$\mu$$\times$$\sigma$$^{2}_{\alpha}$$\times$
ln($\lambda$$_{\rm X}$/$\lambda$$_{o}$)}.

\noindent
In this expression $\mu$ is the exponent in the population law 
N(S)$\sim$S$^{-\mu}$ giving the number of sources with flux densities in excess
of a specified value S (Williams \& Bridle \cite{williams67}). With $\mu$=1.87, 
$\lambda$$_{o}$=5\,000 \AA\ and $\lambda$$_{\rm X}$=12.4 \AA\ (1 keV), we find 
that to get the observed increased value of F$_{X}$/F$_{V}$ by a factor 25, 
corresponding to {$\alpha$$_{\rm X}$--$\alpha$$_{o}$=0.54, a value of 
$\sigma$$_{\alpha}$=0.22 is required, {\it i.e.} $\sigma$=0.57, where 
$\sigma$=2.60$\times$$\sigma_{\alpha}$ is the standard deviation of 
log(F$_{X}$/F$_{V}$). This is in reasonable agreement with our estimated value 
of $\sigma$$\sim$0.5. \\

 Table \ref{xopt} confirms that radio loud QSOs are brighter X-ray sources than 
radio quiet QSOs. In addition we found that, for BL~Lac objects, the mean 
F$_{X}$/F$_{V}$ ratio is larger than that of the radio loud QSOs as previously 
shown by Stocke et al. (\cite{stocke85}). BL~Lac brighter than 15.5 are all 
detected by ROSAT while at O$_{\rm APS}$=20.0, 50\% are still  detected. 

\subsection{The surface density of bright QSOs}

 The surface density of bright QSOs is difficult to determine because these objects 
are rare and therefore large areas of the sky have to be surveyed to find a substantial 
number. The Bright QSO Survey (BQS) of Schmidt \& Green (\cite{schmidt83})
covering 10\,714 deg$^{2}$, contains 69 QSOs brighter than M$_{\rm B}$=--24.0 and 
B=16.16, corresponding to 0.0064$\pm$0.0008 deg$^{-2}$. 
 A number of works suggest that this value is substantially underestimated and that 
the completeness of the BQS is in the range 30\%-70\% (Wampler \& Ponz \cite{wampler85}; 
Goldschmidt et al. \cite{goldschmidt92}; La Franca \& Cristiani \cite{franca97}; 
Wisotzki et al. \cite{wisotzki00}; Mickaelian et al. \cite{mickaelian01}). 

 Grazian et al. (\cite{grazian02}) have undertaken a program similar to ours. 
They cross-correlated the ROSAT-BSC with the Digitized Sky Survey in the southern 
hemisphere to find all coincidences between X-ray sources and stellar objects brighter 
than B$_{\rm J}$$<$15.13. After spectroscopic observations of the candidates,
they obtained a sample of 111 AGN in an area of 5\,660 deg$^{2}$, 57  being 
brighter than M$_{\rm B}$=--24.0. But the distribution of absolute magnitudes 
is rather odd, 46 being greater than --25.0 and 5 smaller than --26.0.
We checked the appearance of some of the low redshift objects on the DSS images; 
they are clearly galaxies; their O magnitude in the APS data base is much greater
than the B magnitude given by Grazian et al. and the absolute magnitudes computed 
with these O magnitudes are definitely greater than --24.0. The true number of QSOs 
in this sample is therefore unknown, but certainly significantly smaller than 57. \\

 There are, in our region of interest (10\,313 deg$^{2}$), 503 known QSOs 
(M$_{\rm B}$$<$--24.0 ${\it i.e.}$ M$_{\rm O}$$<$--24.4) brighter than O$_{\rm USNO}$=16.5, 
${\it i.e.}$ B=16.9 (on the average, the O$_{\rm USNO}$ magnitudes are brighter than 
the photoelectric B magnitudes: $<$O$_{\rm USNO}$--B$>$=--0.38; Mickaelian et al. 
\cite{mickaelian01}). 191 are associated with a BSC source, 84 with a FSC source and 
228 are not detected by ROSAT. Here we use USNO rather than APS O magnitudes because the
selection has been made with the USNO data base; using the APS magnitudes, our 
sample would not be complete to O=16.5. 

  These 503 QSOs correspond to a surface density of 0.041$\pm$0.002
QSO deg$^{-2}$ brighter than B=16.9. If the integrated number-magnitude relation 
has a slope equal to b=$\mu$/2.5=0.75 (Sandage \& Luyten \cite{sandage69}), the 
surface density of QSOs brighter than B=16.16 is 0.0114 deg$^{-2}$. Of these QSOs 
142 are brighter than O$_{\rm USNO}$=15.8 (B=16.2), corresponding to 0.014$\pm$0.001 
deg$^{-2}$, in agreement with the preceding value, confirming that the value 
of b used is reasonable. 

 The QSO counts are systematically affected by the photometric errors on the 
magnitudes as these errors scatter many more objects toward bright magnitudes 
than it does toward faint magnitudes. Assuming that the error distribution 
is Gaussian, with a dispersion $\sigma$, the correction to be applied to the 
observed counts is a factor 10$^{(b+1)\sigma^{2}/2}$ (Eddington 
\cite{eddington40}). With b=0.75 and $\sigma$=0.31, the true QSO surface density 
is smaller by 1.21 than the observed one or, at B=16.16, 0.011$\pm$0.002 deg$^{-2}$. 
This is already 78\% larger than the original Schmidt \& Green estimate.

  Sixty seven of the known QSOs (M$_{\rm O}$$<$--24.4) with available APS magnitudes 
are associated with a ROSAT-FSC source, 35 being located in zone IVa (Table 
\ref{xrayq}). In this zone, we have found 33 additional such objects (see Table 
\ref{catalog}), almost doubling the number of known QSOs at this limiting magnitude. 

  We should note in addition that, among the 84 EBL-WK objects, 14 have not yet been 
spectroscopically observed, while 5 of the QSOs candidates in Table \ref{catalog} 
have still to be observed. This may increases the number of new bright QSOs.
  
\section{Conclusion}

 We have undertaken a program of systematic identification of optically bright AGN
in both the BSC- and FSC-ROSAT catalogues. In this paper we present preliminary 
results concerning the FSC catalogue. We describe the method used which consists
of first cross-correlating the USNO and APS databases with the X-ray catalogue, then
checking the nature of the candidates by looking at the slitless Hamburg spectra.

 Among the starlike objects brighter than B=16.9 we have recovered 35 previously 
known QSOs in the region of the O/(O--E) diagram containing $60\%$ of the bluest 
QSOs and discovered 33 new ones confirming that, even at these relatively bright 
magnitudes, available surveys are significantly incomplete.

  The total number of previously known QSOs brighter than B=16.2 corresponds to a
surface density 78\% higher than the original Schmidt \& Green value. But we have 
shown that, in the subsample studied here ${\it i.e.}$ the QSOs associated with a 
ROSAT-FSC source, the completeness is at best 50\%. Although we cannot generalize 
this to the whole sample, it is not unlikely that the true surface density is 
significantly higher than the number derived here. 

 We found that at this magnitude and at a flux limit of 0.02 counts s$^{-1}$, only 
40\% of all QSOs are detected by ROSAT.
  
\small
\section{Acknowledgments}

 RM and VCh acknowledge CONACyT research grants J32178-E and 39560-F. This work is 
partly based on observations obtained with 2.6-m telescope at the Byurakan 
Astrophysical Observatory, Armenia, the 1.93-m telescope at the Observatoire de 
Haute Provence (CNRS), France and the 2.1-m at the Guillermo Haro Astrophysical 
Observatory, Cananea, Mexico. The authors acknowledge the hospitality of INAOE 
during the Guillermo Haro Workshop 2003, at which time the scientific conclusions 
presented here were discussed. They are grateful to Dr. J. Cabanela for his help 
with the use of the APS databases which are supported by the National Science 
Foundation, the National Aeronautics and Space Administration, and the University 
of Minnesota, and are available at http://aps.umn.edu/. The Digitized Sky Survey 
was produced at the Space Telescope Science Institute (STScI) under U.S. Government 
grant NAG W-2166.

\end{document}